\documentclass[11pt,onecolumn,draftclsnofoot]{IEEEtran}
\usepackage{amsfonts}
\IEEEoverridecommandlockouts

\ifCLASSINFOpdf
\else
\fi
\usepackage{epsfig}
\usepackage{graphicx}
\usepackage{psfig}
\usepackage{epsf}
\usepackage[cmex10]{amsmath}
\usepackage{booktabs}
\usepackage{fancyhdr}
\usepackage{slashbox}
\usepackage{caption2}   
\newtheorem{myTheo}{Theorem}   

\newtheorem{rem}{Remark} 
\begin{document}
\title{Robust AN-Aided Beamforming and Power Splitting Design for Secure MISO Cognitive Radio With SWIPT}
\author{\IEEEauthorblockN{Fuhui Zhou, \emph{Student Member, IEEE}, Zan Li, \emph{Senior Member, IEEE},\\ Julian Cheng, \emph{Senior Member, IEEE}, Qunwei Li, and Jiangbo Si}
\thanks{F. Zhou, Z. Li and J. Si are with the Integrated Service Networks Lab of
Xidian University, Xi’an, 710071, China (e-mail: zhoufuhui1989@163.com, \{zanli, jbsi\}@xidian.edu.cn).

J. Cheng is with the School of Engineering, The University of British Columbia, Kelowna, BC, Canada (e-mail: julian.cheng@ubc.ca).

Q. Li is with the Department of Electrical Engineering and Computer Science, Syracuse University, Syracuse, NY 13244 USA (e-mail: lee880716@gmail.com).
}

\thanks{The research was supported by the Natural Science Foundation of China (61301179, 61501356, 61501354 and 61401338) and a scholarship from China Scholarship Council.}}
\maketitle
\begin{abstract}
A multiple-input single-output cognitive radio downlink network is studied with simultaneous wireless information and power transfer. In this network, a secondary user coexists with multiple primary users and multiple energy harvesting receivers. In order to guarantee secure communication and energy harvesting, the problem of robust secure artificial noise-aided beamforming and power splitting design is investigated under imperfect channel state information (CSI). Specifically, the transmit power minimization problem and the max-min fairness energy harvesting problem are formulated for both the bounded CSI error model and the probabilistic CSI error model. These problems are non-convex and challenging to solve. A one-dimensional search algorithm is proposed to solve these problems based on ${\cal S}\text{-Procedure} $ under the bounded CSI error model and based on Bernstein-type inequalities under the probabilistic CSI error model. It is shown that the optimal robust secure beamforming can be achieved under the bounded CSI error model, whereas a suboptimal beamforming solution can be obtained under the probabilistic CSI error model. A tradeoff is elucidated between the secrecy rate of the secondary user receiver and the energy harvested by the energy harvesting receivers under a max-min fairness criterion.
\end{abstract}
\IEEEpeerreviewmaketitle
\section{Introduction}
\IEEEPARstart{T}{HE} unprecedented increase of mobile devices and escalating high data rate requirements have resulted in severe spectrum scarcity problem. Cognitive radio (CR) is a promising technique that aims to utilize spectrum efficiently and alleviate the spectrum scarcity problem \cite{Haykin}. In CR under spectrum sharing, a secondary user (SU) can coexist with a primary user (PU) based on the condition that the interference caused by the SU is tolerable to the PU. Since CR has the potential to improve spectrum efficiency, it has been widely investigated for traditional cellular networks, relay networks, and wireless sensor networks \cite{O. Akan} and \cite{X. Huang}. On the other hand, the escalating requirement for high data rate and the ubiquitous wireless services have also contributed to the sharp growth of energy consumption and resulted in energy scarcity problem, especially for CR with energy-constrained devices such as limited-energy wireless sensors and cellular phones \cite{X. Huang}-\cite{C. Jiang}. In order to address the energy scarcity problem, a number of works focused on maximizing the energy efficiency of CR networks \cite{X. Huang}-\cite{A. Alabbasi}.

Recently, a promising technology called simultaneous wireless information and power transfer (SWIPT) has been proposed to solve the energy scarcity problem, and it has attracted much attention \cite{Y. Dong}-\cite{X. Lu}. Specifically, radio frequency (RF) signals radiated by transmitters have dual purposes. RF signals not only can carry information, but also can be used as a source for wireless power transfer charging the energy-constrained communication devices. Thus, in CR with energy-constrained communication devices such as cellular phones, it is of great importance to investigate CR with SWIPT that can improve spectrum efficiency and energy utilization simultaneously \cite{Y. Chen}-\cite{D. W. K. Ng1}. However, due to the inherent characteristics of CR with SWIPT, malicious energy harvesting receivers (EHRs) may exist and illegitimately access the PU bands or change the radio environment. As a result, the legitimate SU is unable to use frequency bands of the PU or has his confidential transmitted information intercepted \cite{D. W. K. Ng1} and \cite{B. Fang}. Thus, the security of CR with SWIPT is also of crucial importance.

Physical-layer security, which is based on the physical layer characteristics of the wireless channels, has been proposed to improve the security of wireless communication systems \cite{R. Liu}. However, it was shown that the secrecy rate of a wireless communication system with physical-layer security is limited by the channel state information (CSI). In CR, the secrecy rate of the SU is further limited since the transmit power of the SU should be controlled to protect the PU from harmful interference \cite{Y. Pei}. In order to improve the secrecy rate of the SU, both multi-antenna technique and beamforming technique have been introduced \cite{Y. Pei}-\cite{C. Wang}. However, the optimal beamforming schemes proposed in \cite{Y. Pei}-\cite{C. Wang} may not be appropriate in CR with SWIPT since energy harvested by EHRs should be considered. On the other hand, in practice, it is difficult to obtain the perfect CSI due to the existence of channel estimation errors and quantization errors, especially in CR with SWIPT, where there is no cooperation among SUs, PUs and EHRs. Even worse, the imperfect CSI can significantly deteriorate the beamforming performance. Thus, it is of great importance to design robust secure beamforming for CR with SWIPT.

Based on the model of CSI errors, there are two different approaches to designing robust secure beamforming. The first approach uses a bounded set to model the CSI errors. It has the advantage in implementation complexity but the obtained results can underestimate the actual performance. The second approach uses a probabilistic model to describe the CSI error. In this case, it finds solutions that can be efficiently computed and provide a good approximation to outage-based constraints. The second approach is mainly applicable to scenarios where there exist delay-sensitive communication devices or where constraints on the secure beamforming design are loose. These two approaches have been widely applied to design robust secure beamforming for the traditional wireless communication systems \cite{J. Huang}-\cite{Q. Li3}, wireless communication systems with SWIPT \cite{D. W. K. Ng2}-\cite{F. Wang}, and CR with SWIPT \cite{D. W. K. Ng1}. The works in \cite{J. Huang}-\cite{F. Wang} are related to this paper and they are summarized as follows.

\textbf{Traditional wireless communication systems:} The design of robust secure beamforming has been analyzed under the bounded CSI error model in \cite{J. Huang}-\cite{Q. Li2}, and under the probabilistic CSI error model in \cite{S. Ma2}-\cite{Q. Li3}. Specifically, using the worst-case optimization, robust secure transmission schemes were studied for multiple-input single-output (MISO) wiretap channels in \cite{J. Huang}, and for multiple-input multiple-output (MIMO) wiretap channels in \cite{Z. Chu}. It was shown that by introducing a cooperative jammer, the achievable secrecy rate can be significantly improved. In \cite{Q. Li1} and \cite{Q. Li2}, the authors extended the design of robust secure transmission schemes to the scenarios where there exist several multi-antenna eavesdroppers. An artificial noise (AN)-aided transmit strategy was used in \cite{Q. Li2}. It was shown that the AN-aided transmit strategy is an efficient way to improve the secrecy rate. In order to improve the performance under the worst-case scenario, i.e., the bounded CSI error model, robust secure beamforming was designed under outage-based constraints for MISO wiretap channels \cite{S. Ma2}-\cite{Z. Chu2}. In \cite{K. Y. Wang}, three approximate forms for the outage-based constraint were derived, and this work was extended to the scenario with several multi-antenna eavesdroppers \cite{Z. Chu2}. Using the Bernstein-type inequality \cite{Q. Li4}, \cite{Q. Li3}, the robust secure beamforming was designed for MISO wiretap channels with an AN-aided transmit strategy \cite{Q. Li4} and for MIMO wiretap channels \cite{Q. Li3}. However, the robust secure beamforming schemes proposed in \cite{J. Huang}-\cite{Q. Li3} are not appropriate in wireless communication systems with SWIPT, because the energy harvesting requirement of EHRs is not considered.

\textbf{Wireless communication systems with SWIPT:} In order to guarantee secure communication and energy harvesting requirement, robust secure beamforming design problems were studied in wireless communication systems with SWIPT \cite{D. W. K. Ng2}-\cite{F. Wang}. Using the bounded CSI error model, robust secure beamforming schemes were designed to optimize different objectives over the MISO secrecy channels, such as the transmit power \cite{D. W. K. Ng2}, \cite{D. W. K. Ng3}, the minimum energy harvested by EHRs \cite{M. R. A. Khandaker}, the secrecy rate and energy harvested by EHRs \cite{Z. Chu3}. The work in \cite{D. W. K. Ng2}, \cite{D. W. K. Ng3} was extended to the MIMO system \cite{S. H. Wang}. In \cite{D. W. K. Ng2}-\cite{S. H. Wang},  the bounded CSI error model was used, whereas the probabilistic CSI error model was used for the MISO system \cite{F. Wang}.

\subsection{Motivation and Contributions}
Compared with the traditional wireless communication systems with SWIPT, a robust secure beamforming design in secure CR with SWIPT is more important and meaningful. In secure CR with SWIPT, the robust design of beamforming not only provides the SU with a reasonably high secrecy rate, but also protects the PU from harmful interference. Moreover, in secure CR with SWIPT, the robust beamforming design can guarantee energy harvesting requirement and prolong the operation time. Although the robust beamforming design issue has been investigated in the traditional wireless communication systems with SWIPT in \cite{D. W. K. Ng2}-\cite{S. H. Wang}, few study, with the exception of \cite{D. W. K. Ng1}, has been devoted to designing robust beamforming schemes for CR with SWIPT.

In this paper, we focus on designing robust secure beamforming schemes for MISO CR with SWIPT under both the bounded CSI error model and the probabilistic CSI error model. The optimization objectives are the minimum transmit power of the cognitive base station (CBS) or the energy harvested by EHRs under a max-min fairness criterion. The system model considered in our work is pertaining to that in \cite{D. W. K. Ng1}. However, our work differs from \cite{D. W. K. Ng1} in four aspects. Firstly, in our paper, a power splitting receiver architecture is adopted at the SU receiver. However, it has not been considered in  \cite{D. W. K. Ng1} and thus the desired SU receiver cannot simultaneously decode information and harvest energy.  Moreover, we jointly optimize the power splitting ratio in order to further improve the achievable performance. Secondly, the interference power leakage-to-transmit power ratio was identified as the optimization objective in \cite{D. W. K. Ng1}. In this case, the interference imposed on PUs from the SU may be intolerable. In our work, in order to protect PUs from harmful interference, interference power constraints are imposed. Thirdly, different from \cite{D. W. K. Ng1}, we focus on designing robust secure beamforming to minimize the transmit power of the CBS or to maximize the energy harvested by EHRs under a max-min fairness criterion. Finally, both the bounded CSI error model and the probabilistic CSI error model are considered here, whereas only the bounded case was considered in \cite{D. W. K. Ng1}. We now summarize the main contributions of this paper as follows.

\begin{enumerate}
  \item Robust beamforming design problems are studied for secure MISO CR with SWIPT under the bounded CSI error model. Robust AN-aided beamforming and the power splitting ratio are jointly designed, and they are subject to the constraints on the secrecy rate, the interference power, and the transmit power. The optimization objectives are the minimum transmit power of the CBS or the energy harvested by EHRs under a max-min fairness criterion. These two problems are non-convex and challenging, because there are infinite inequality constraints and coupling among different variables. In order to solve those two problems, a one-dimensional search algorithm is proposed based on the semidefinite relaxation (SDR) and ${\cal S}$-Procedure \cite{A. Ben-Tal}, \cite{S. P. Boyd}. It is proved that the optimal robust secure beamforming can always be found and the rank of the optimal AN covariance matrix is unity.
  \item Both an outage-constrained transmit power minimization problem and an outage EH maximization problem are formulated in secure MISO CR with SWIPT under the probabilistic CSI error model. We jointly design robust AN-aided beamforming and the power splitting ratio under the outage secrecy rate constraint, the maximum interference probabilistic constraint, and the outage EH constraint. Since there are no closed-form expressions for these probabilistic constraints, approximations to these constraints, which are based on the Bernstein-type inequalities \cite{A. Ben-Tal}, \cite{I. Bechar}, are made in order to make the formulated problems tractable. It is shown that the optimal robust secure beamforming cannot always be obtained and a suboptimal beamforming solution is found.
  \item Simulation results show that robust secure beamforming under the probabilistic CSI error model can provide performance gains compared with that under the bounded CSI error model at the cost of higher implementation complexity. A tradeoff is found between the secrecy rate of the SU receiver and the energy harvested by EHRs under a max-min fairness criterion.
\end{enumerate}

\subsection{Organization and Notations}
The remainder of this paper is organized as follows. Section II presents the system model. Robust secure beamforming design problems under the bounded CSI error model are examined in Section III. Section IV presents robust secure beamforming design problems under the probabilistic CSI error model. Simulation results are presented in Section V. Finally, Section VI concludes the paper.

\emph{Notations:} Matrices and vectors are denoted by boldface capital letters and boldface lower case letters, respectively. $\mathbf{I}$ denotes the identity matrix; vec(\textbf{A}) denotes the vectorization of matrix \textbf{A} and it is obtained by stacking its column vectors. The Hermitian (conjugate) transpose, trace, and rank of a matrix \textbf{A} are denoted respectively by $\mathbf{A^H}$, Tr$\left(\mathbf{A}\right)$ and Rank$\left(\mathbf{A}\right)$. $\mathbf{x}^\dag$ represents the conjugate transpose of a vector $\mathbf{x}$. $\mathbf{C}^{M\times N}$ stands for a $M$-by-$N$ dimensional complex matrix set. $\mathbf{A}\succeq \mathbf{0} \left(\mathbf{A}\succ \mathbf{0}\right)$ represents that $\mathbf{A}$ is a Hermitian positive semidefinite (definite) matrix. $\mathbb{H}^N$ and $\mathbb{H}_+^{N}$ represent a $N$-by-$N$ dimensional Hermitian matrix set and a Hermitian positive semidefinite matrix set, respectively. ${\left\|  \cdot  \right\|}$ denotes the Euclidean norm of a vector. ${\left| \cdot \right|}$ represents the absolute value of a complex scalar. $\mathbf{x} \sim {\cal C}{\cal N}\left( {\mathbf{u},\mathbf{\Sigma } }\right)$ means that $\mathbf{x}$ is a random vector, which follows a complex Gaussian distribution with mean $\mathbf{u}$ and covariance matrix $\mathbf{\Sigma }$. $\mathbb{E}[ \cdot ]$ denotes the expectation operator. ${\rm Re}\left(\mathbf{a}\right)$ extracts the real part of vector $\mathbf{a}$. $\mathbb{R}_{+}$ represents the set of all nonnegative real numbers.

\section{System Model}
\begin{figure}[!t]
\centering
\includegraphics[width=5.0 in]{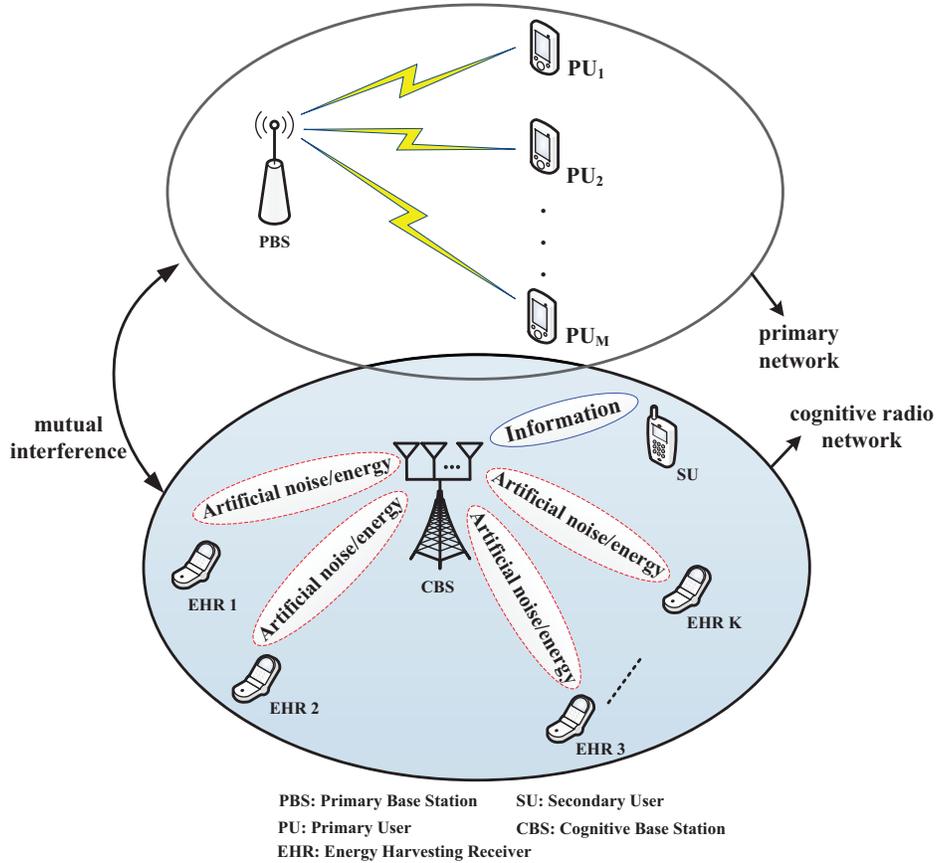}
\caption{The system model.} \label{fig.1}
\end{figure}
A downlink MISO CR network with SWIPT under spectrum sharing is considered in Fig. 1, where one secondary link, $M$ primary links, and $K$ energy harvesting links share the same spectrum. The CBS is equipped with $N_t$ antennas. The primary base station (PBS), the SU receiver, each PU receiver and each EHR are equipped with single antenna. The PBS transmits information to $M$ PU receivers while the CBS provides a SWIPT service to the SU receiver, and transfers energy to $K$ EHRs as long as the interference imposed on $M$ PU receivers from the CBS is tolerable. It is assumed that the SU receiver is an energy-constrained device such as a cellular phone or a wireless sensor whose battery has limited energy storage capacity \cite{I. Krikidis}-\cite{B. Fang}. Different from \cite{D. W. K. Ng1}, a power splitting receiver architecture is adopted in the SU receiver. Thus, the SU receiver splits the received signal power into two parts in order to simultaneously decode information and harvest energy. Due to the inherent characteristics of CR and SWIPT, EHRs may eavesdrop and intercept the information transmitted by the CBS. In this paper, we assume that PU receivers are friendly users and do not eavesdrop the information sent by the CBS. This assumption has been widely used in secure CR \cite{I. Krikidis}, \cite{X. Lu}, \cite{Y. Pei}-\cite{C. Wang}, \cite{R. K. Sharma}-\cite{N. Mokari}. All the channels involved are assumed to be slow frequency-nonselective fading channels. Let ${\cal K}$ and ${\cal I}$ denote the set ${\cal K} \buildrel \Delta \over = \left\{ {1,2, \cdots ,K} \right\}$ and the set ${\cal I} \buildrel \Delta \over = \left\{ {1,2, \cdots ,M} \right\}$, respectively. The signals received at the SU receiver, the $k$th EHR, and the interference signals imposed on the $i$th PU receiver from the CBS, denoted by $y$, ${y_{e,k}}$, and ${y_{PU,i}}$, where $k \in \cal K$ and $i \in \cal I$, can be given, respectively, as
\begin{subequations}
\begin{align}\label{27}\
&y = {\mathbf{h}^\dag }\mathbf{x} + {n_s}\\
&{y_{e,k}} = \mathbf{g}_k^\dag \mathbf{x} + {n_{e,k}},\ k \in \cal K\\
&{y_{PU,i}} = \mathbf{q}_i^\dag \mathbf{x},\ i \in \cal I
\end{align}
\end{subequations}
where $\mathbf{h} \in {\mathbf{C}^{{N_t} \times 1}}$, $\mathbf{g}_k \in {\mathbf{C}^{{N_t} \times 1}}$ and $\mathbf{q}_k \in {\mathbf{C}^{{N_t} \times 1}}$ are the channel vectors between the CBS and the SU receiver, the $k$th EHR and the $i$th PU receiver, respectively. In $\left(1\right)$, ${n_s} \sim {\cal C}{\cal N}\left( {0,\sigma _s^2} \right)$ and ${n_{e,k}} \sim {\cal C}{\cal N}\left( {0,\sigma _e^2} \right)$ respectively denote the complex Gaussian noise at the SU receiver and the $k$th EHR, which include the interference from the PBS and the additive complex Gaussian noise at the SU receiver and the $k$th EHR. Note that the interferences at the SU receiver and the $k$th EHR from the PBS are assumed to be circularly symmetric complex Gaussian. This interference model is the worst case model and has been widely used in \cite{D. W. K. Ng1}, \cite{R. K. Sharma}-\cite{N. Mokari}. This model also covers the system models studied in \cite{B. Fang}, \cite{Y. Pei}-\cite{L. Zhang}, where the interferences from the PBS to the SU receiver and the $k$th EHR are negligible. In $\left(1\right)$, $\mathbf{x} \in {\mathbf{C}^{{N_t} \times 1}}$ is the transmit signal vector given as
\begin{align}\label{27}\
\mathbf{x} = \mathbf{w}s + \mathbf{v}
\end{align}
where $s \in {\mathbf{C}^{{1} \times 1}}$ and $\mathbf{w}\in {\mathbf{C}^{{N_t} \times 1}}$ denote the confidential information-bearing signal for the SU receiver and the corresponding beamforming vector, respectively. Without loss of generality, it is assumed that $\mathbb{E}[ {{{\left| s \right|}^2}} ] = 1$. $\mathbf{v}\in {\mathbf{C}^{{N_t} \times 1}}$ is the noise vector artificially generated by the CBS in order to improve the secrecy rate of the SU and energy transfer at EHRs. It is assumed that $\mathbf{v} \sim {\cal C}{\cal N}\left( {0,\mathbf{\Sigma} } \right)$, where $\mathbf{\Sigma}$ is the AN covariance matrix to be designed. Since the SU receiver is equipped with a power splitting device, the equivalent signal received at the SU receiver for information decoding, denoted by ${y_D}$, is given as
\begin{align}\label{27}\
{y_D} = \sqrt \rho  \left( {{ \mathbf{h}^\dag } \mathbf{x} + {n_s}} \right) + {n_{s,p}}
\end{align}
where ${n_{s,p}}$ is the additional processing noise generated by the SU receiver. It is assumed that ${n_{s,p}} \sim {\cal C}{\cal N}\left( {0,\sigma _{s,p}^2} \right)$ and ${n_{s,p}}$ is independent of ${n_s}$ and ${n_{e,k}}$. The parameter $\rho  \in \left( {0,1} \right)$ denotes the power splitting ratio, which denotes that $\rho$ portion of the received power is used to decode information and $1-\rho$ portion of the received power is used to transfer energy. Thus, the secrecy rate of the SU receiver, denoted by $R_s$, is given as
\begin{subequations}
\begin{align}\label{27}\
&{R_s} ={\mathop {\min }\limits_{k \in \cal K} }\ {\left\{ { C_s- C_{e,k}} \right\}}\\
& C_s={{\log }_2}\left( {1 + \frac{{\rho {\mathbf{h}^\dag}\mathbf{w}{\mathbf{w}^\dag}\mathbf{h}}}{{\rho \left( {{\mathbf{h}^\dag}\mathbf{\Sigma} \mathbf{h} + \sigma _s^2} \right) + \sigma _{s,p}^2}}} \right)\\
& C_{e,k}={{\log }_2}\left( {1 + \frac{{\mathbf{g}_k^\dag\mathbf{w}{\mathbf{w}^\dag}{\mathbf{g}_k}}}{{\mathbf{g}_k^\dag\mathbf{\Sigma} {\mathbf{g}_k} + \sigma _e^2}}} \right)
\end{align}
\end{subequations}
and the energy harvested at the SU receiver and the $k$th EHR, denoted by $E_s$ and $E_{e,k}$, can be respectively expressed as
\begin{subequations}
\begin{align}\label{27}\
&{E_s} = \left( {1 - \rho } \right)\eta \left( {{\mathbf{h}^\dag }\mathbf{w}{\mathbf{w}^\dag }\mathbf{h} + {\mathbf{h}^\dag }\mathbf{\Sigma} \mathbf{h} + \sigma _s^2} \right)\\
&{E_{e,k}} = \eta \left( {\mathbf{g}_k^\dag \mathbf{w}{\mathbf{w}^\dag }{\mathbf{g}_k} + \mathbf{g}_k^\dag \mathbf{\Sigma} {\mathbf{g}_k} + \sigma _e^2} \right),\ k \in \cal K
\end{align}
\end{subequations}
where $\eta  \in \left( {0,\left. 1 \right]} \right.$ is a constant that denotes the energy conversion efficiency at the SU receiver and the EHRs.

It is assumed that the CSI $\mathbf{h}$ is available at both CBS and SU \cite{D. W. K. Ng1}, \cite{Z. Chu2}-\cite{Z. Chu3}. However, the channel vectors $\mathbf{g}_k, k \in \cal K$ and $\mathbf{q}_i, i \in \cal I$ cannot be perfectly known since there is no cooperation among the CBS, PU receivers and EHRs \cite{D. W. K. Ng1}, \cite{Z. Chu2}-\cite{Z. Chu3}.
\section{Worst-Case Robust Secure Beamforming Design}
In this section, the robust secure beamforming design problems are considered under the bounded CSI error model. The transmit power of the CBS is minimized and the energy harvested by the EHRs under a max-min fairness criterion is maximized subject to the secrecy rate constraint, energy harvesting constraints, interference power constraints, and the transmit power constraint. In order to solve the robust secure beamforming design problems, which are non-convex and have infinite inequality constraints, a one-dimensional search algorithm is proposed based on SDR and ${\cal S}$-Procedure. The inner algorithm solves a convex optimization problem, which can be solved by using the software \texttt{CVX} \cite{M. Grant}.
\subsection{The Bounded CSI Error Model}
Quantization errors for CSI can be modeled as the bounded CSI error forms, which are deterministic models \cite{S. Ma}, \cite{J. Huang}-\cite{Q. Li2}. The bounded CSI error model for the channel vector $\mathbf{g}_k, k \in \cal K$, is given as
\begin{subequations}
\begin{align}\label{27}\
&{\mathbf{g}_k} = \overline {{\mathbf{g}_k}}  + \Delta {\mathbf{g}_k},\ k \in \cal K\\
&{\mathbf{\Psi} _{e,k}} \buildrel \Delta \over = \left\{ {\Delta {\mathbf{g}_k} \in {\mathbf{C}^{{N_t} \times 1}}:\Delta \mathbf{g}_k^\dag \Delta {\mathbf{g}_k} \le \xi _{e,k}^2} \right\}
\end{align}
\end{subequations}
and the bounded CSI error model for the channel vector $\mathbf{q}_i, i \in \cal I$, is given as
\begin{subequations}
\begin{align}\label{27}\
&{\mathbf{q}_i} = \overline {{\mathbf{q}_i}}  + \Delta {\mathbf{q}_i},\ i \in \cal I\\
&{\mathbf{\Psi} _{P,i}} \buildrel \Delta \over = \left\{ {\Delta {\mathbf{q}_i} \in {\mathbf{C}^{{N_t} \times 1}}:\Delta \mathbf{q}_i^\dag \Delta {\mathbf{q}_i} \le \xi _{P,i}^2} \right\}
\end{align}
\end{subequations}
where $\overline {{\mathbf{g}_k}}$ and $\overline {{\mathbf{q}_i}}$ are the estimates of the channel vectors $\mathbf{g}_k$ and $\mathbf{q}_m$, respectively; ${\mathbf{\Psi} _{e,k}} $ and ${\mathbf{\Psi} _{P,i}}$ denote the uncertainty regions of $\mathbf{g}_k$ and $\mathbf{q}_m$; $\Delta {\mathbf{g}_k}$ and $\Delta {\mathbf{q}_i}$ represent the channel estimation errors of $\mathbf{g}_k$ and $\mathbf{q}_m$; $\xi _{e,k}$ and $\xi _{P,i}$ are the radiuses of the uncertainty regions ${\mathbf{\Psi} _{e,k}} $ and ${\mathbf{\Psi} _{P,i}}$, respectively.

\subsection{The Worst-Case Transmit Power Minimization Problem}
Let $\mathbf{W}$ and $\mathbf{H}$ denote $\mathbf{W} = \mathbf{w}{\mathbf{w}^\dag }$ and $\mathbf{H} = \mathbf{h}{\mathbf{h}^\dag }$, respectively. Based on the bounded CSI error model, the worst-case transmit power minimization problem subject to the secrecy rate constraint, EH constraints, interference power constraints and the transmit power constraint, denoted by $\textbf{P}_{{1}}$, is given as
\begin{subequations}
\begin{align}\label{27}\
\textbf{P}_{{1}}: \ \ \ \ \ &{\mathop {\min }\limits_{\mathbf{W},\mathbf{\Sigma} ,\rho } }\ {\text{Tr}\left( {\mathbf{W} + \mathbf{\Sigma} } \right)} \\
&\text{s.t.}\ \ \  C1:{R_s} \ge R_{\min},\ \forall \Delta {\mathbf{g}_k} \in {\mathbf{\Psi} _{e,k}},\ k \in \cal K\\
&\ \ \ \ \ \ \  C2:\left( {1 - \rho } \right)\eta \left( {\mathbf{Tr}\left( {\mathbf{W}\mathbf{H} + \mathbf{\Sigma} \mathbf{H}} \right) + \sigma _s^2} \right) \ge {\psi _s}\\
&\ \ \ \ \ \ \  C3:\eta \left( {\mathbf{g}_k^\dag \left( {\mathbf{W} + \mathbf{\Sigma} } \right){\mathbf{g}_k} + \sigma _e^2} \right) \ge {\psi _{e,k}},\ \forall \Delta {\mathbf{g}_k} \in {\mathbf{\Psi} _{e,k}}\\
&\ \ \ \ \ \ \  C4:\mathbf{q}_i^\dag \left( {\mathbf{W} + \mathbf{\Sigma} } \right){\mathbf{q}_i} \le {P_{In,i}},\ \forall \Delta {\mathbf{q}_i} \in {\mathbf{\Psi} _{P,i}},\ i \in \cal I\\
&\ \ \ \ \ \ \  C5:\text{Tr}\left( {\mathbf{W} + \mathbf{\Sigma} } \right) \le {P_{th}}\\
&\ \ \ \ \ \ \  C6:0 < \rho  < 1\\
&\ \ \ \ \ \ \  C7:\text{Rank}\left( \mathbf{W} \right) = 1\\
&\ \ \ \ \ \ \  C8:\mathbf{\Sigma} \succeq0, \ \mathbf{W}\succeq0
\end{align}
\end{subequations}
where $R_{\min}$ is the minimum secrecy rate requirement of the CBS; ${\psi _s}$ and ${\psi _{e,k}}$ denote the required minimum harvesting energy required at the SU receiver and the $k$th EHR, respectively; ${P_{In,i}}$ represents the maximum tolerable interference power of the $i$th PU receiver; $P_{th}$ is the maximum transmit power of the CBS. In $\left(8\rm{b}\right)$-$\left(8\rm{i}\right)$, $C1$ can guarantee that the secrecy rate of the SU is not less than $R_{\min}$; $C2$ and $C3$ are the EH constraints of the SU receiver and EHRs; the interference power constraint $C4$ is imposed in order to protect the QoS of PUs; $C5$ limits the maximum transmit power of the CBS. Note that $\textbf{P}_{{1}}$ may be infeasible due to the transmit power constraint $C5$. Owing to the non-convex constraints in $C2$ and $C7$, $\textbf{P}_{{1}}$ is non-convex and difficult to be solved. Moreover, there are infinite inequality constraints to be satisfied due to the uncertain regions, ${\mathbf{\Psi} _{e,k}} $ and ${\mathbf{\Psi} _{P,i}}$, which make $\textbf{P}_{{1}}$ even more challenging. In order to solve $\textbf{P}_{{1}}$, SDR and ${\cal S}$-Procedure are applied as follows.

\emph{\textbf{Lemma 1 $\left({\cal S}\text{-Procedure} \right) $ }} \cite{S. P. Boyd}: Let $ {f_i}\left( \mathbf{z} \right) = {\mathbf{z}^\dag }{\mathbf{A}_i}\mathbf{z} + 2{\mathop{\mathbf{\rm Re}}\nolimits} \left\{ {\mathbf{b}_i^\dag \mathbf{z}} \right\} + {c_i}, i \in \left\{ {1,2} \right\}$, where $\mathbf{z}\in \mathbf{C}^{N\times 1}$, $\mathbf{A}_i \in \mathbb{H}^N$, $\mathbf{b}_i\in \mathbf{C}^{N\times 1}$ and $c_i\in \mathbb{R}$. Then, the expression ${f_1}\left( \mathbf{z} \right) \le 0 \Rightarrow {f_2}\left( \mathbf{z} \right) \le 0$ holds if and only if there exists a $\alpha  \ge 0$ such that
\begin{align}\label{27}\
\alpha \left[ {\begin{array}{*{20}{c}}
{{\mathbf{A}_1}}&{{\mathbf{b}_1}}\\
{\mathbf{b}_1^\dag }&{{c_1}}
\end{array}} \right] - \left[ {\begin{array}{*{20}{c}}
{{\mathbf{A}_2}}&{{\mathbf{b}_2}}\\
{\mathbf{b}_2^\dag }&{{c_2}}
\end{array}} \right]
\succeq \mathbf{0}
\end{align}
provided that there exists a vector $\mathbf{\widehat z}$ such that ${f_i}\left( {\mathbf{\widehat z}} \right) < 0$.

By introducing slack variables, $t=1/\rho$, and $\beta$, where $t>1$ and $\beta\geq1$, and applying \textbf{Lemma 1}, the secrecy rate constraint $C1$ can be equivalently expressed as
\begin{subequations}
\begin{align}\label{27}\
&\mathbf{Tr}\left\{ {\left( {\mathbf{W} + \left( {1 - {2^{{R_{\min }}}}\beta } \right)\mathbf{\Sigma} } \right)\mathbf{H}} \right\} + \left( {1 - {2^{{R_{\min }}}}\beta } \right)\left( {\sigma _s^2 + \sigma _{s,p}^2t} \right) \ge 0\\ \notag
&{\mathbf{\Gamma} _k}\left( {{\omega _k},\mathbf{W},\mathbf{\Sigma} ,\beta } \right)\\ &=\left[ {\begin{array}{*{20}{c}}
{{\omega _k}\mathbf{I} - \left( {\mathbf{W} - \left( {\beta  - 1} \right)\mathbf{\Sigma} } \right)}&{ - \left( {\mathbf{W} - \left( {\beta  - 1} \right)\mathbf{\Sigma} } \right)\overline {{\mathbf{g}_k}} }\\
{ - {{\overline {{\mathbf{g}_k}} }^\dag }\left( {\mathbf{W} - \left( {\beta  - 1} \right)\mathbf{\Sigma} } \right)}&{\left( {\beta  - 1} \right)\sigma _e^2 - {{\overline {{\mathbf{g}_k}} }^\dag }\left( {W - \left( {\beta  - 1} \right)\mathbf{\Sigma} } \right)\overline {{\mathbf{g}_k}}  - {\omega _k}\xi _{e,k}^2}
\end{array}} \right]\succeq\mathbf{0} ,\ k\in \cal K
\end{align}
\end{subequations}
where ${\omega _k} \ge 0, k\in \cal K$, is a slack variable. The proof of $\left(10\right)$ can be found in Appendix A.
Similarly, by applying \textbf{Lemma 1}, the constraints $C3$ and $C4$ can be equivalently expressed as
\begin{subequations}
\begin{align}\label{27}\
&{\mathbf{\Gamma} _k}\left( {{\mu _k},\mathbf{W},\mathbf{\Sigma} } \right)=\left[ {\begin{array}{*{20}{c}}
{{\mu _k}\mathbf{I} + \left( {\mathbf{W} + \mathbf{\Sigma} } \right)}&{\left( {\mathbf{W} + \mathbf{\Sigma} } \right)\Delta {\mathbf{g}_k}}\\
{{{\overline {{\mathbf{g}_k}} }^\dag }\left( {\mathbf{W} + \mathbf{\Sigma} } \right)}&{{{\overline {{\mathbf{g}_k}} }^\dag }\left( {\mathbf{W} + \mathbf{\Sigma} } \right)\overline {{\mathbf{g}_k}}  + \sigma _e^2 - {\psi _{e,k}}{\eta ^{ - 1}} - {\mu _k}\xi _{e,k}^2}
\end{array}} \right]\succeq\mathbf{0}, \ \ k\in \cal K
\\
&{\mathbf{\Gamma} _i}\left( {{\delta _i},\mathbf{W},\mathbf{\Sigma} } \right)=\left[ {\begin{array}{*{20}{c}}
{{\delta _i}\mathbf{I} - \left( {\mathbf{W} + \mathbf{\Sigma} } \right)}&{ - \left( {\mathbf{W} + \mathbf{\Sigma} } \right)\overline {{\mathbf{q}_i}} }\\
{ - {{\overline {{\mathbf{q}_i}} }^\dag }\left( {\mathbf{W} + \mathbf{\Sigma} } \right)}&{{P_{In,i}} - {{\overline {{\mathbf{q}_i}} }^\dag }\left( {\mathbf{W} + \mathbf{\Sigma} } \right)\overline {{\mathbf{q}_i}}  - {\delta _i}\xi _{P,i}^2}
\end{array}} \right]\succeq\mathbf{0},\ \ i\in \cal I
\end{align}
\end{subequations}
where $\mu _k\geq0$ and $\delta _i\geq0$, $k\in \cal{K}$, $i\in\cal{I}$, are slack variables. Equation $\left(11\right)$ can be similarly obtained by following Appendix A. By using SDR, $\textbf{P}_{{1}}$ can be relaxed to the following problem as
\begin{subequations}
\begin{align}\label{27}\
\textbf{P}_{{2}}: \ \ \ \ \ &{\mathop {\min }\limits_{\mathbf{W},\mathbf{\Sigma} ,t,\beta, \{\omega _k\},\{\mu _k\},\{\delta _i\} } }\ {\mathbf{Tr}\left( {\mathbf{W} + \mathbf{\Sigma} } \right)} \\
&\text{s.t.}\ \ \  \mathbf{Tr}\left\{ {\left( {\mathbf{W} + \mathbf{\Sigma} } \right)\mathbf{H}} \right\} + \sigma _s^2 - \frac{{{\psi _s}}}{\eta }\left( {1 + \frac{1}{{t - 1}}} \right) \ge 0\\
&\ \ \ \ \ \ \  t  > 1\\
&\ \ \ \ \ \ \  \beta\geq1\\
&\ \ \ \ \ \ \  \omega _k, \mu _k, \delta _i\geq0\\
&\ \ \ \ \ \ \ \left(10\right), \left(11\right), C5, C8.
\end{align}
\end{subequations}
Unfortunately, since the variable $\beta$ couples with the other variables, $\textbf{P}_{{2}}$ is still non-convex. Using the fact that $R_{\min}\geq0$, $\left(8\rm{b}\right)$ and $\left(8\rm{f}\right)$, one can obtain $\beta  \le 1 + {P_{th}}{{\left\| \mathbf{h} \right\|}^2}/\sigma _{s,p}^2$. Thus, the following inequalities can be obtained
\begin{align}\label{27}\
1 \le \beta  \le 1 + \frac{{{P_{th}}}}{{\sigma _{s,p}^2}}{\left\| \mathbf{h} \right\|^2}.
\end{align}
While $\textbf{P}_{{2}}$ is non-convex, it is however convex for a given $\beta$. Thus, the optimal $\beta$ value can be searched from the interval $[ {1,1 + {P_{th}}{{\left\| \mathbf{h} \right\|}^2}/\sigma _{s,p}^2}]$, such that the optimal values of $\mathbf{W}$, $\mathbf{\Sigma}$, and $\beta$ will achieve the minimum transmit power in $\textbf{P}_{{2}}$. Using the uniform sampling method, \textbf{Algorithm 1} is proposed to solve $\textbf{P}_{{2}}$. Table I summarizes the details of \textbf{Algorithm 1}. We can now state the following two theorems.
\begin{table}[htbp]
\begin{center}
\caption{The one-dimensional line search algorithm}
\begin{tabular}{lcl}
\\\toprule
$\textbf{Algorithm 1}$: The one-dimensional line search algorithm\\ \midrule
\  1: \textbf{Setting:}\\
\ \ \ \ \ \ \ the minimum secrecy rate $R_{\min}$, the minimum EH, ${\psi _s}$ and ${\psi _{e,k}},\ k \in \cal K$, \\
\ \ \ \ \ \ \ the maximum tolerable interference power ${P_{In,i}},\ i\in\cal I$, the maximum transmit power,$P_{th}$,\\
\ \ \ \ \ \ \ the radiuses of the uncertainty regions, $\xi _{e,k}$, $\xi _{P,i}$ and the uniform search step, $\tau_1$.\\
\  2: \textbf{Inputting:}\\
\ \ \ \ \ \ \ the CSI,$\mathbf{h}$, the estimated CSI, $\overline {{\mathbf{g}_k}}$ and $\overline {{\mathbf{q}_i}}$.\\
\  3: \textbf{Initialization:}\\
\ \ \ \ \ \ \ the iteration index $n=1$. \\
\  4: \textbf{Optimization:}\\
\ 5:\ \ \ \ \textbf{$\pmb{\unrhd} $  for  $\beta$=1:$\tau_1$:$1 + {P_{th}}{{\left\| \mathbf{h} \right\|}^2}/\sigma _{s,p}^2$}\\
\ 6:\ \ \ \ \ \ \ \  \ use the software \texttt{CVX} to solve the optimization problem $\textbf{P}_{{3}}$ for the given $\beta$, given as \\
\ \ \ \ \ \ \ \ \ \ \ \  \ $ \textbf{P}_{{3}}: \ \ \ \ \ {\mathop {\min }\limits_{\mathbf{W},\mathbf{\Sigma} ,t, \{\omega _k\},\{\mu _k\},\{\delta _i\} } }\ {\mathbf{Tr}\left( {\mathbf{W} + \mathbf{\Sigma} } \right)} $\\
\ \ \ \ \ \ \ \ \ \ \ \ \ \ \ \ \ \ \ \  $ \text{s.t.} \ \ \ \left(10\right), \left(11\right), \left(12\rm{b}\right), \left(12\rm{c}\right), \left(12\rm{e}\right), {C_5}\ \text{and}\ {C_8}\ \text{are satisfied}.$ \\
\ 7:\ \ \ \ \ \ \ \ obtain the robust matrix, $\mathbf{W}^n$, $\mathbf{\Sigma}^n$, the slack variable, $t^n$, and $\mathbf{Tr}\left( {\mathbf{W}^n + \mathbf{\Sigma}^n } \right)$;\\
\ 8: \ \ \ \ \ \ \ set $n=n+1$;\\
\ 9:\ \ \ \ \textbf{$\pmb{\unrhd} $ end} \\
\  10: \textbf{Comparison:}\\
\ 11:\ \ \ \ \  \  compare all $\mathbf{Tr}\left( {\mathbf{W}^n + \mathbf{\Sigma}^n } \right)$ and obtain the minimum $\mathbf{Tr}\left( {\mathbf{W} + \mathbf{\Sigma} } \right)$, \\
\ \ \ \ \ \ \ \ \ \  \  the optimal robust matrices, $\mathbf{W}^{opt}$, $\mathbf{\Sigma}^{opt}$ and optimal $\rho^{opt}=1/t^{opt}$.\\
\  12: \textbf{Eigenvalue decomposition:}\\
\ 13:\ \ \ \ \  \  perform eigenvalue decomposition for $\mathbf{W}^{opt}$ and obtain the optimal robust beamforming $\mathbf{w}^{opt}$.\\
\bottomrule
\end{tabular}
\end{center}
\end{table}

\begin{myTheo}
For problem $\textbf{P}_{{2}}$, assume that the minimum secrecy rate $R_{\min}>0$ and that $\textbf{P}_{{2}}$ is feasible, the optimal $\mathbf{W}$ is unique and its rank is one.
\end{myTheo}
\begin{IEEEproof}
See Appendix B.
\end{IEEEproof}
\begin{rem}
Theorem 1 indicates that the transmit power minimization given as $\textbf{P}_{{1}}$ can be obtained by solving the rank-relaxed problem presented in $\textbf{P}_{{2}}$. The optimal transmit matrix $\mathbf{W}$ is rank-one in spite of the numbers of the PU receivers and EHRs. Moreover, for any $\beta$, the optimal $\mathbf{W}$ is rank-one. Thus, the optimal robust secure beamforming vector can be obtained by eigenvalue decomposition over $\mathbf{W}$, which is the eigenvector related to the maximum eigenvalue of $\mathbf{W}$.
\end{rem}
\begin{myTheo}
The AN covariance matrix of $\textbf{P}_{{2}}$, $\mathbf{\Sigma}$, has the rank-one property regardless of the number of PU receivers and the number of the EHRs.
\end{myTheo}
\begin{IEEEproof}
The proof is similar to the proof for Theorem 1 and it is omitted due to space limitation.
\end{IEEEproof}
\subsection{The Worst-Case Max-Min Fairness EH Problem}
In this subsection, in order to provide fairness to EHRs under the bounded CSI error model, we consider the robust max-min fairness EH problem. The robust secure beamforming design for the max-min fairness EH, under the secrecy rate constraint, the EH constraint of the SU receiver, interference power constraints and the transmit power constraint, can be formulated as the following problem
\begin{subequations}
\begin{align}\label{27}\
\textbf{P}_{{4}}: \ \ \ \ \ &{\mathop {\max }\limits_{\mathbf{W},\mathbf{\Sigma} } } \ {\mathop {\min }\limits_{\Delta {\mathbf{g}_k} \in {\mathbf{\Psi }_{e,k}}, k \in \cal{K}} } {\eta \left( {\mathbf{g}_k^\dag\left( {\mathbf{W} + \mathbf{\Sigma} } \right){\mathbf{g}_k} + \sigma _e^2} \right)}\\
&\text{s.t.}\ \ \  C1, C2, C4-C8.
\end{align}
\end{subequations}
Note that $\textbf{P}_{{4}}$ may be infeasible due to the stated constraints. Using a slack variable, $\tau$, we can express $\textbf{P}_{{4}}$ as the following equivalent problem
\begin{subequations}
\begin{align}\label{27}\
\textbf{P}_{{5}}: \ \ \ \ \ &{\mathop {\max }\limits_{\mathbf{W},\mathbf{\Sigma} } }\  \tau \\
&\text{s.t.}\ \ \ \eta \left( {\mathbf{g}_k^\dag \left( {\mathbf{W} + \mathbf{\Sigma} } \right){\mathbf{g}_k} + \sigma _e^2} \right) \ge \tau ,\Delta {\mathbf{g}_k} \in {\mathbf{\Psi } _{e,k}},k \in \cal K \\
& \ \ \ \ \ \ \ C1, C2, C4-C8.
\end{align}
\end{subequations}
By applying \textbf{Lemma 1}, the constraint $\left(15\rm{b}\right)$ can be equivalently written as
\begin{align}\label{27}\
{\mathbf{\Gamma} _k}\left( {{\mu _k},\mathbf{W},\mathbf{\Sigma}, \tau } \right)=\left[ {\begin{array}{*{20}{c}}
{{\mu _k}\mathbf{I} + \left( {\mathbf{W} + \mathbf{\Sigma} } \right)}&{\left( {\mathbf{W} + \mathbf{\Sigma} } \right)\Delta {\mathbf{g}_k}}\\
{{{\overline {{\mathbf{g}_k}} }^\dag }\left( {\mathbf{W} + \mathbf{\Sigma} } \right)}&{{{\overline {{\mathbf{g}_k}} }^\dag }\left( {\mathbf{W} + \mathbf{\Sigma} } \right)\overline {{\mathbf{g}_k}}  + \sigma _e^2 - \tau{\eta ^{ - 1}} - {\mu _k}\xi _{e,k}^2}
\end{array}} \right]\succeq\mathbf{0}, \ \ k\in \cal K
\end{align}
where $\mu _k\geq0$ is a slack variable. Following a similar procedure, $\textbf{P}_{{5}}$ can be relaxed as
\begin{subequations}
\begin{align}\label{27}\
\textbf{P}_{{6}}: \ \ \ \ \ &{\mathop {\max }\limits_{\mathbf{W},\mathbf{\Sigma} ,t,\beta, \{\omega _k\},\{\mu _k\},\{\delta _i\} } }\  \tau \\
&\text{s.t.}\ \ \ \left(10\right), \left(11\rm{b}\right), \left(12\rm{b}\right)-\left(12\rm{e}\right), \left(16\right), C5\  \text{and}\ C8
\end{align}
\end{subequations}
where $\omega _k$ and $\delta _i$ are slack variables associated with the secrecy rate constraint and the interference power constraint, $C1$ and $C4$, respectively. Problem $\textbf{P}_{{6}}$ is still non-convex due to the coupling between $\beta$ and $\mathbf{\Sigma}$ in $\left(10\rm{b}\right)$. Similar to $\textbf{P}_{{2}}$, it is seen that $\textbf{P}_{{6}}$ is convex for a given $\beta$. Thus, \textbf{Algorithm 1} can be modified to solve $\textbf{P}_{{6}}$. In this case, the minimum EH constraint of EHRs is dropped and $\textbf{P}_{{3}}$ is replaced by the following problem
\begin{subequations}
\begin{align}\label{27}\
\textbf{P}_{{7}}: \ \ \ \ \ &{\mathop {\max }\limits_{\mathbf{W},\mathbf{\Sigma} ,t, \{\omega _k\},\{\mu _k\},\{\delta _i\} } }\  \tau \\
&\text{s.t.}\ \ \ \left(17\rm{b}\right).
\end{align}
\end{subequations}
By solving $\textbf{P}_{{6}}$, Theorem 3 and Theorem 4 can be stated as follows.
\begin{myTheo}
 Assuming that the minimum secrecy rate $R_{\min}>0$ and that $\textbf{P}_{{6}}$ is feasible, the optimal $\mathbf{W}$ of $\textbf{P}_{{6}}$ is unique and has rank one.
\end{myTheo}

\begin{IEEEproof}
The proof is similar to the proof of Theorem 1, and it is based on the Karush-Kuhn-Tucker (KKT) optimality conditions of $\textbf{P}_{{6}}$. This proof is omitted due to space limitation.
\end{IEEEproof}
\begin{rem}
Theorem 3 implies that the optimal robust secure beamforming for $\textbf{P}_{{4}}$ can be obtained by solving the relaxed problem presented in $\textbf{P}_{{6}}$. The solution obtained by solving $\textbf{P}_{{6}}$ is optimal regardless of the numbers of PU receivers and EHRs.
\end{rem}
\begin{myTheo}
The optimal AN covariance matrix of $\textbf{P}_{{6}}$ is rank-one if $R_{\min}>0$ and $\textbf{P}_{{6}}$ is feasible.
\end{myTheo}
\begin{IEEEproof}
The proof is similar to the proof for Theorem 1 and it is omitted due to space limitation.
\end{IEEEproof}
\begin{rem}
Since the step 6 of \textbf{Algorithm 1} solves a convex optimization problem, which can be solved by using, for example, the interior-point method \cite{S. P. Boyd}. According to \cite{A. Ben-Tal1}, the complexity of \textbf{Algorithm 1} is due to three parts, namely, the uniform line search, iteration complexity and the per-iteration computation cost. Since the number of the positive semidefinite matrix constraints for $\textbf{P}_{{3}}$ is the same as that of $\textbf{P}_{{7}}$, the difference between the complexity of $\textbf{P}_{{3}}$ and that of $\textbf{P}_{{7}}$ lies in the number of the uniform search. It is assumed that the accuracy of the iteration is $\varepsilon$ and the number of the uniform search is $T_1$. Then, the total complexity of \textbf{Algorithm 1} under the bounded CSI error model is given as \begin{align}\label{27}\ \notag
&{T_1}\ln \left( {{\varepsilon ^{ - 1}}} \right){\cal O}\left( {KMN_t^2} \right)\sqrt {\left( {2K + M + 2} \right){N_t} + 2M + 4K + 4} \left\{ {2N_t^2\left( {{N_t} + {\cal O}\left( {KMN_t^2} \right)} \right)} \right.\\ \notag
&+ \left( {4 + 2K + M} \right)\left( {1 + {\cal O}\left( {KMN_t^2} \right)} \right) + \left( {2K + M} \right)\left[ {{{\left( {{N_t} + 1} \right)}^3} + {\cal O}\left( {KMN_t^2} \right){{\left( {{N_t} + 1} \right)}^2}} \right]\\
&\left. { + {{\cal O}^2}\left( {KMN_t^2} \right)} \right\}
\end{align}
\end{rem}
where ${\cal O}\left( {\cdot} \right)$ is the big-$\rm O$ notation.
\section{Outage-Constrained Robust Secure Beamforming Design}
The robust secure beamforming design problems under the bounded CSI error model can guarantee that the SU achieves the worst-case performance. However, the results obtained under the bounded CSI error model are conservative since the extreme scenario may happen rarely \cite{S. Ma}, \cite{S. Ma2}-\cite{Q. Li3}. Moreover, robust secure beamforming designed under the bounded error model is inappropriate to the delay-sensitive CR where user applications such as voice and video can be sensitive to delay. Alternatively, the outage-constrained robust secure beamformig design is more suitable, and it is designed under the probabilistic CSI errors model \cite{S. Ma}, \cite{S. Ma2}-\cite{Q. Li3}. Thus, in this section, the robust secure beamforming design problems are studied under the probabilistic constraints.
\subsection{The Probabilistic CSI Error Model}
In a probabilistic CSI error model, the CSI error is stochastic and follows a certain distribution, instead of within a determined region. Specifically, the complex circular Gaussian CSI error model is applied, which has been widely used \cite{S. Ma}, \cite{S. Ma2}-\cite{Q. Li3}. In this case, the channel vectors, $\mathbf{g}_k, k \in \cal K$, and $\mathbf{q}_i, i \in \cal I$, can be given as
\begin{subequations}
\begin{align}\label{27}\
&{\mathbf{g}_k} = \overline {{\mathbf{g}_k}}  + \Delta {\mathbf{g}_k}, \ \Delta {\mathbf{g}_k} \sim {\cal C}{\cal N}\left( {0,{\mathbf{G}_k}} \right), \ k \in \cal K \\
&{\mathbf{q}_i} = \overline {{\mathbf{q}_i}}  + \Delta {\mathbf{q}_i},\ \Delta {\mathbf{q}_i} \sim {\cal C}{\cal N}\left( {0,{\mathbf{Q}_i}} \right) \ i \in \cal I
\end{align}
\end{subequations}
where $\overline {{\mathbf{g}_k}}$ and $\overline {{\mathbf{q}_i}}$ are the estimated CSI, which are known at the CBS; $\Delta {\mathbf{g}_k}$ and $\Delta {\mathbf{q}_i}$ denote the channel error vectors, respectively; $\mathbf{G}_k$ and $\mathbf{Q}_i$ are the covariance matrices of the corresponding channel error vectors; $\Delta {\mathbf{g}_k}$ and $\Delta {\mathbf{g}_j}$ are independent for any $k\neq j$; $\Delta {\mathbf{q}_i}$ is independent of $\Delta {\mathbf{q}_j}$ for any $i\neq j$.

\subsection{The Outage-Constrained Transmit Power Minimization Problem}
In this subsection, based on the probabilistic CSI error model, the transmit power minimization problem is considered under the probabilistic constraints. In this case, the transmit power minimization problem is formulated as follows
\begin{subequations}
\begin{align}\label{27}\
\textbf{P}_{{8}}: \ \ \ \ &{\mathop {\min }\limits_{\mathbf{W},\mathbf{\Sigma} ,\rho } }\ {\text{Tr}\left( {\mathbf{W} + \mathbf{\Sigma} } \right)} \\
&\text{s.t.}\  \ \ \Pr\left\{ C_s-C_{e,k}\geq R_{\min} \right\} \ge 1 - {\varpi _r},\ k \in \cal K\\
&\ \ \ \ \ \   \Pr\left\{ {\eta \left( {\mathbf{g}_k^\dag \left( {\mathbf{W} + \mathbf{\Sigma} } \right){\mathbf{g}_k} + \sigma _e^2} \right) \ge {\psi _{e,k}}} \right\} \ge 1 - {\varpi _{e,k}},k \in \cal K\\
&\ \ \ \ \ \   \Pr\left\{ {\mathbf{q}_i^\dag \left( {\mathbf{W} + \mathbf{\Sigma} } \right){\mathbf{q}_i} \le {P_{In,i}}} \right\} \ge 1 - {\varpi _{I,i}},i \in \cal I\\
&\ \ \ \ \ \   C2, C5-C8
\end{align}
\end{subequations}
where $\varpi _r\in (0,1]$, $\varpi _{e,k}\in (0,1]$ and $\varpi _{I,i}\in (0,1]$ denote the maximum outage probabilities associated with the secrecy rate and the EH of $k$th EHR, and the maximum tolerable probability that the interference power caused to the $i$th PU is larger than ${P_{In,i}}$, respectively. The outage secrecy rate constraint presented by $\left(21\rm{b}\right)$ guarantees that the probability of the secrecy rate being larger than $R_{\min}$ is larger than $1 - {\varpi _r}$. The constraint given by $\left(21\rm{c}\right)$ is to guarantee that the outage probability of the minimum EH requirement of the $k$th EHR is less than ${\varpi _{e,k}}$. The constraint presented by $\left(21\rm{c}\right)$ guarantees that the probability that the interference power causing to the $i$th PU is larger than $P_{In,i}$ is less than ${\varpi _{I,i}}$. It is difficult to solve $\textbf{P}_{{8}}$ since $\textbf{P}_{{8}}$ is non-convex and the expressions given by $\left(21\rm{b}\right)$, $\left(21\rm{c}\right)$ and $\left(21\rm{d}\right)$ have no closed-form expressions. In order to solve $\textbf{P}_{{8}}$, a safe approximation to $\textbf{P}_{{8}}$ is given based on Bernstein-type inequalities. Here, \lq\lq safe\rq\rq \ means that the obtained solution based on such an approximation always satisfies the outage-based constraints \cite{Q. Li3}.

\emph{\textbf{Lemma 2} }\emph{The Bernstein-type Inequality I}  \cite{K. Y. Wang}: Let $f\left( \mathbf{z} \right) = {\mathbf{z}^\dag }\mathbf{A}\mathbf{z} + 2{\mathop{\rm Re}\nolimits} \left\{ {{\mathbf{z}^\dag }\mathbf{b}} \right\} + c$, where $\mathbf{A} \in \mathbb{H}^N$, $\mathbf{b}\in \mathbf{C}^{N\times 1}$, $c\in \mathbb{R}$ and $z\sim {\cal C}{\cal N}\left( {\mathbf{0},{\mathbf{I}}} \right)$. For any $\varpi\in (0 \ 1]$, an approximate and convex form for
\begin{align}\label{27}\
\Pr \left\{ {f\left( \mathbf{z} \right) \ge 0} \right\} \ge 1 - \varpi
\end{align}
can be written as
\begin{subequations}
\begin{align}\label{27}\
&\text{Tr}\left( \mathbf{A} \right) - \sqrt { - 2\ln \left( \varpi  \right)} {\upsilon _1} + \ln \left( \varpi  \right){\upsilon _2} + c \ge 0 \\
&\left\| {\left[ \begin{array}{l}
\text{vec}\left( \mathbf{A} \right)\\
\sqrt 2 \mathbf{b}
\end{array} \right]} \right\| \le {\upsilon _1}
\\
&{\upsilon _2}\mathbf{I} +\mathbf{A}\succeq \mathbf{0},{\upsilon _2} \ge 0
\end{align}
\end{subequations}
where $\upsilon _1$ and $\upsilon _2$ are slack variables.

\emph{\textbf{Lemma 3} }\emph{The Bernstein-type Inequality II}  \cite{{I. Bechar}}: Let $f\left( \mathbf{z} \right) = {\mathbf{z}^\dag }\mathbf{A}\mathbf{z} + 2{\mathop{\rm Re}\nolimits} \left\{ {{\mathbf{z}^\dag }\mathbf{b}} \right\} + c$, where $\mathbf{A} \in \mathbb{H}^N$, $\mathbf{b}\in \mathbf{C}^{N\times 1}$, $c\in \mathbb{R}$ and $z\sim {\cal C}{\cal N}\left( {\mathbf{0},{\mathbf{I}}} \right)$. For any $\varpi\in (0\ 1]$, an approximate and convex form for the constraint
\begin{align}\label{27}\
\Pr \left\{ {f\left( \mathbf{z} \right) \leq 0} \right\} \ge 1 - \varpi
\end{align}
can be written as
\begin{subequations}
\begin{align}\label{27}\
&\text{Tr}\left( \mathbf{A} \right) + \sqrt { - 2\ln \left( \varpi  \right)} {\upsilon _1} - \ln \left( \varpi  \right){\upsilon _2} + c \le 0 \\
&\left\| {\left[ \begin{array}{l}
\text{vec}\left(\mathbf{A} \right)\\
\sqrt 2 \mathbf{b}
\end{array} \right]} \right\| \le {\upsilon _1}
\\
&{\upsilon _2}\mathbf{I} - \mathbf{A}\succeq \mathbf{0},\ {\upsilon _2} \ge 0
\end{align}
\end{subequations}
where $\upsilon _1$ and $\upsilon _2$ are slack variables.

Let $\Delta {\mathbf{g}_k} = \mathbf{G}_k^{1/2}{\mathbf{\widehat g}_k}$, where ${\mathbf{\widehat g}_k} \sim {\cal C}{\cal N}\left( {0,{\mathbf{I}}} \right)$ and $k\in \cal K$. By applying \textbf{Lemma 2}, an approximation to the outage secrecy rate constraint presented in $\left(21\rm{b}\right)$ can be given as
\begin{subequations}
\begin{align}\label{27}\
&z = \frac{{\rho \left( {{\mathbf{h}^\dag }\mathbf{\Sigma} \mathbf{h} + \sigma _s^2} \right) + \sigma _{s,p}^2}}{{\rho \left( {{\mathbf{h}^\dag }\left( {\mathbf{\Sigma } + \mathbf{W}} \right)\mathbf{h} + \sigma _s^2} \right) + \sigma _{s,p}^2}} \\
&\text{Tr}\left( { \mathbf{G}_k^{1/2}\mathbf{M}\mathbf{G}_k^{1/2}} \right) - \sqrt { - 2\ln {\varpi _r}} {\delta _{r,k}} + \ln {\varpi _r}{\nu _{r,k}} + {c_{r,k}} \ge 0,\  k\in \cal K\\
&{c_{r,k}} = {\overline {{g_k}} ^\dag }\mathbf{M}\overline {{g_k}}  - \left( {{2^{{R_{\min }}}}z - 1} \right)\sigma _e^2,\ k\in \cal K\\
&{\left\|\left[ \begin{array}{l}
\text{vec}\left( {\mathbf{G}_k^{1/2}\mathbf{M}\mathbf{G}_k^{1/2}} \right)\\
\sqrt 2 \mathbf{G}_k^{1/2}\mathbf{M}\overline {{\mathbf{g}_k}}
\end{array} \right]\right\|}\le {\delta _{r,k}}, k\in \cal K
\\
&{\nu _{r,k}}\mathbf{I} + \mathbf{G}_k^{1/2}\mathbf{M}\mathbf{G}_k^{1/2}\succeq \mathbf{0},\ {\nu _{r,k}} \ge 0, k\in \cal K
\end{align}
\end{subequations}
where ${\rm \mathbf{M}} = \left( {\left( {1 - {2^R}z} \right)\mathbf{\Sigma}  - {2^R}z\mathbf{W}} \right)$; ${\delta _{r,k}}$ and ${\nu _{r,k}}$ are slack variables. Similar to the outage secrecy rate constraint, using \textbf{Lemma 2}, the outage EH constraint given by $\left(21\rm{c}\right)$ can be approximated as
\begin{subequations}
\begin{align}\label{27}\
&\text{Tr}\left( {\mathbf{G}_k^{1/2}\left( {\mathbf{W} + \mathbf{\Sigma} } \right)\mathbf{G}_k^{1/2}} \right) - \sqrt { - 2\ln \left( {\varpi _{e,k}} \right)} {\delta _{e,k}} + \ln \left( {{\varpi _{e,k}}} \right){\nu _{e,k}} + {c_{e,k}} \ge 0, \  k\in \cal K\\
&{c_{e,k}} = \sigma _e^2 + {\overline {{\mathbf{g}_k}} ^\dag}\left( {\mathbf{W} + \mathbf{\Sigma} } \right)\overline {{\mathbf{g}_k}}  - \frac{{{\psi _{e,k}}}}{\eta }, \  k\in \cal K\\
&\left\| {\left[ \begin{array}{l}
\text{vec}\left( {\mathbf{G}_k^{1/2}\left( {\mathbf{W} + \mathbf{\Sigma} } \right)\mathbf{G}_k^{1/2}} \right)\\
\sqrt 2 \mathbf{G}_k^{1/2}\left( {\mathbf{W} + \mathbf{\Sigma} } \right)\overline {{\mathbf{g}_k}}
\end{array} \right]} \right\| \le {\delta _{e,k}}, \  k\in \cal K
\\
&{\nu _{e,k}}\mathbf{I} + \mathbf{G}_k^{1/2}\left( {\mathbf{W} + \mathbf{\Sigma} } \right)\mathbf{G}_k^{1/2}\succeq  \mathbf{0},\ {\nu _{e,k}} \ge 0, k\in \cal K
\end{align}
\end{subequations}
where $\delta _{e,k}$ and $\nu _{e,k}$ are slack variables. Let $\Delta {\mathbf{q}_i} = \mathbf{Q}_i^{1/2}{\mathbf{\widehat q}_i}$, where ${\mathbf{\widehat q}_i} \sim {\cal C}{\cal N} \left( {0,\mathbf{I}} \right)$ and $i\in \cal I$. Applying \textbf{Lemma 3} to the constraint presented in $\left(21\rm{d}\right)$, an approximation form for this constraint can be given as
\begin{subequations}
\begin{align}\label{27}\
&\text{Tr}\left( {\mathbf{Q}_i^{1/2}\left( {\mathbf{W} + \mathbf{\Sigma} } \right)\mathbf{Q}_i^{1/2}} \right) + \sqrt { - 2\ln {\varpi _{I,i}}} {\delta _{I,i}} - \ln \left({\varpi _{I,i}}\right){\nu _{I,i}} + {c_{q,i}} \le 0, \  i\in \cal I\\
&{c_{q,i}} = \overline {{\mathbf{q}_i}}^\dag \left( {\mathbf{W} + \mathbf{\Sigma} } \right)\overline {{\mathbf{q}_i}}  - {P_{In,i}}, \   i\in \cal I\\
&\left\| {\left[ \begin{array}{l}
\text{vec}\left( {\mathbf{Q}_i^{1/2}\left( {\mathbf{W} + \mathbf{\Sigma} } \right)\mathbf{Q}_i^{1/2}} \right)\\
\sqrt 2 \mathbf{Q}_i^{1/2}\left( {\mathbf{W} + \mathbf{\Sigma} } \right)\overline {{\mathbf{q}_i}}
\end{array} \right]} \right\| \le {\delta _{I,i}}
, \  i\in \cal I
\\
&{\nu _{I,i}}\mathbf{I} - \mathbf{Q}_i^{1/2}\left( {\mathbf{W} + \mathbf{\Sigma} } \right)\mathbf{Q}_i^{1/2} \succeq \mathbf{0},\ {\nu _{I,i}} \ge 0, i\in \cal I
\end{align}
\end{subequations}
where ${\delta _{I,i}}$ and $\nu _{I,i}$ are slack variables. Although the safe approximations to the probabilistic constraints have been applied, the approximation to $\textbf{P}_{{8}}$ is still non-convex since there is a rank-one constraint, a fractional constraint and coupling among different variables. By using SDR, the approximation problem for  $\textbf{P}_{{8}}$ can be relaxed as
\begin{subequations}
\begin{align}\label{27}\
\textbf{P}_{{9}}: \ \ \ \ &{\mathop {\min }\limits_{\mathbf{W},\mathbf{\Sigma} ,\rho,z, \left\{{\delta _{r,k}}\right\} , \left\{{\nu _{r,k}}\right\},\left\{{\delta _{e,k}}\right\} , \left\{{\nu _{e,k}}\right\}, \left\{{\delta _{I,i}}\right\} , \left\{{\nu _{I,i}}\right\}} }\ {\text{Tr}\left( {\mathbf{W} + \mathbf{\Sigma} } \right)} \\
&\text{s.t.}\ \  \left(26\right), \left(27\right), \left(28\right), C2, C5, C6\ \text{and}\ C8.
\end{align}
\end{subequations}
Note that $\textbf{P}_{{9}}$ is still non-convex due to the factional constraint and the coupling of $z$ with the other variables. It is seen from $\left(26\rm{a}\right)$ that $0<z\leq1$. Thus, the optimal $z$ can be obtained by using a one-dimensional line search method. Let $t=1/\rho$. For a given $z$, an approximation to the relaxed $\textbf{P}_{{9}}$ can be given as
\begin{subequations}
\begin{align}\label{27}\
\textbf{P}_{{10}}: \ \ \ \ &{\mathop {\min }\limits_{\mathbf{W},\mathbf{\Sigma} ,t, \left\{{\delta _{r,k}}\right\} , \left\{{\nu _{r,k}}\right\},\left\{{\delta _{e,k}}\right\} , \left\{{\nu _{e,k}}\right\}, \left\{{\delta _{I,i}}\right\} , \left\{{\nu _{I,i}}\right\}} }\ {\mathbf{Tr}\left( {\mathbf{W} + \mathbf{\Sigma} } \right)} \\
&\text{s.t.}\  \ \ {\mathbf{h}^\dag }\left( {\left( {z - 1} \right)\mathbf{\Sigma}  + z\mathbf{W}} \right)\mathbf{h} + \left( {z - 1} \right)\left(t\sigma _{s,p}^2 + \sigma _s^2\right) = 0\\
&\ \ \ \ \ \ \ t>1 \\
&\ \ \ \ \ \  \left(12\rm{b}\right), \left(26\rm{b}\right)-\left(26\rm{e}\right), \left(27\right), \left(28\right), C5 \ \text{and}\ C8.
\end{align}
\end{subequations}
It is seen from $\left(30\right)$ that for a given $z$ $\textbf{P}_{{10}}$ is convex and can be solved by the software \texttt{CVX} \cite{M. Grant}. Based on the above-mentioned analysis, $\textbf{P}_{{9}}$ can be solved by applying the modified \textbf{Algorithm 1}. In this case, ${\varpi _r}$, $\varpi _{e,k}$ and $\varpi _{I,i}$ are required to be initialized. The uniform search variable is $z$ instead of $\beta$ and its search interval is $\left(0\ 1\right]$. Problem $\textbf{P}_{{10}}$ is required to be solved instead of $\textbf{P}_{{3}}$.
\begin{rem}
Problem $\textbf{P}_{{9}}$ cannot guarantee that its solution, $\mathbf{W}$, is rank-one. If the solution $\mathbf{W}$ is rank-one, the optimal robust secure beamforming for $\textbf{P}_{{9}}$ can be obtained by the eigenvalue decomposition. If the solution $\mathbf{W}$ is not rank-one, the well-known Gaussian randomization procedure can be applied to obtain the suboptimal robust secure beamforming vector \cite{Z. Q. Luo}.
\end{rem}
\subsection{The Outage-Constrained Max-Min Fairness EH Problem}
In this subsection, the outage max-min fairness EH problem is first formulated subject to the probabilistic constraints, and this problem has not been studied before. The max-min fairness EH problem is formulated under the probabilistic CSI error model as
\begin{subequations}
\begin{align}\label{27}\
\textbf{P}_{{11}}: \ \ \ \ &{\mathop {\max }\limits_{\mathbf{W},\mathbf{\Sigma} ,\rho } }\ \ {{\Gamma _E}}\\
&\text{s.t.}\  \ \ \Pr \left\{ {\eta \left( {\mathbf{g}_k^\dag \left( {\mathbf{W}+ \mathbf{\Sigma} } \right){\mathbf{g}_k} + \sigma _e^2} \right) \ge {\Gamma _E}} \right\} \ge 1 - {\varpi _{e,k}},\ k \in \cal K\\
&\ \ \ \ \ \  \left(21\rm{b}\right), \left(21\rm{d}\right), C2, C5-C8
\end{align}
\end{subequations}
where ${\Gamma _E}$ is the outage energy harvested by the EHRs under a max-min fairness criterion. Note that $\textbf{P}_{{11}}$ can guarantee that the probability of the outage max-min fairness energy harvested by the EHRs satisfies the following inequality
\begin{align}\label{27}\ \notag
\Pr \left\{ {\begin{array}{*{20}{c}}
{\mathop {\min }\limits_{k \in K} }\ {\eta \left( {\mathbf{g}_k^\dag \left( {\mathbf{W}+ \mathbf{\Sigma} } \right){\mathbf{g}_k} + \sigma _e^2} \right) \ge {\Gamma _E}}
\end{array}} \right\}
& = \prod\limits_{k = 1}^K {\Pr \left\{ {\eta \left( {\mathbf{g}_k^\dag \left( {\mathbf{W}+ \mathbf{\Sigma} } \right){\mathbf{g}_k} + \sigma _e^2} \right) \ge {\Gamma _E}} \right\}} \\
 &\ge \prod\limits_{k = 1}^K {\left( {1 - {\varpi _{e,k}}} \right)}.
\end{align}

Problem $\textbf{P}_{{11}}$ is also challenging due to the probabilistic constraints. In order to solve $\textbf{P}_{{11}}$, using the same techniques as those used for solving $\textbf{P}_{{8}}$, a modified \textbf{Algorithm 1} can be used to solve the approximate problem for $\textbf{P}_{{11}}$. For a given $z$, the approximate problem is given as
\begin{subequations}
\begin{align}\label{27}\
\textbf{P}_{{12}}: \ \ \ \ &{\mathop {\max }\limits_{\mathbf{W},\mathbf{\Sigma} ,t, \left\{{\delta _{r,k}}\right\} , \left\{{\nu _{r,k}}\right\},\left\{{\delta _{e,k}}\right\} , \left\{{\nu _{e,k}}\right\}, \left\{{\delta _{I,i}}\right\} , \left\{{\nu _{I,i}}\right\}} }\ \ {{\Gamma _E}}\\
&\text{s.t.}\  \ \ {c_{e,k}} = \sigma _e^2 + {\overline {{\mathbf{g}_k}} ^\dag}\left( {\mathbf{W} + \mathbf{\Sigma} } \right)\overline {{\mathbf{g}_k}}  - \frac{{{\Gamma _E}}}{\eta }, \  k\in \cal K\\
&\ \ \ \ \ \  \left(12\rm{b}\right), \left(26\rm{b}\right)-\left(26\rm{e}\right), \left(27\rm{a}\right), \left(27\rm{c}\right), \left(27\rm{d}\right), \left(28\right), \left(30\rm{b}\right),\left(30\rm{c}\right) ,C5 \ \text{and}\ C8
\end{align}
\end{subequations}
where $\delta _{r,k}\geq0$ and $\nu _{r,k}\geq0$, $\delta _{e,k}\geq0$ and $\nu _{e,k}\geq0$, $\delta _{I,i}\geq0$ and $\nu _{I,i}\geq0$, are the slack variables associated with the outage secrecy rate constraint in $\left(21\rm{b}\right)$, the constraint in $\left(31\rm{b}\right)$ and the constraint in $\left(21\rm{d}\right)$, respectively. $t$ is equal to $1/\rho$. Problem $\textbf{P}_{{12}}$ is convex and can be solved by the software \texttt{CVX}. Note that the solution $\mathbf{W}$ for $\textbf{P}_{{12}}$ may be not rank-one. Thus, the optimal beamforming vector $\mathbf{w}^{opt}$ for $\textbf{P}_{{12}}$ is obtained when $\mathbf{W}$ is rank-one. Otherwise, the suboptimal beamforming vector $\mathbf{w}$ is attained by using the well-known Gaussian randomization procedure.
\begin{rem}
The complexity of \textbf{Algorithm 1} under the probabilistic CSI error model also comes from three parts, namely, the uniform line search, iteration complexity and the per-iteration computation cost. Since the numbers of the norm-constraints and the positive semidefinite matrix constraints for $\textbf{P}_{{10}}$ are the same as those of the relaxed approximation to $\textbf{P}_{{12}}$, the only difference of their complexities lies in the number of uniform searches. Assume that the accuracy of the iteration is $\varepsilon$ and the number of the uniform searches is $T_2$, then the total complexity of the modified \textbf{Algorithm 1} under the probabilistic CSI error model is given by
\begin{align}\label{27}\ \notag
&{T_2}\ln \left( {{\varepsilon ^{ - 1}}} \right){\cal O}\left( {KMN_t^2} \right)\sqrt {\left( {2K + M + 2} \right){N_t} + 3M + 6K + 4}\\ \notag
&\ \ \times\left\{ {\left( {2K + M + 2} \right)N_t^2\left( {{N_t} + {\cal O}\left( {KMN_t^2} \right)} \right) + \left( {2K + M + 4} \right)\left( {1 + {\cal O}\left( {KMN_t^2} \right)} \right)} \right.\\
&\ \ \ \left. { + \left( {2K + M} \right){{\left[ {{N_t}\left( {{N_t} + 1} \right) + 1} \right]}^2} + {{\cal O}^2}\left( {KMN_t^2} \right)} \right\}.
\end{align}
\end{rem}
It is seen from $\left(19\right)$ and $\left(34\right)$ that the effect of ${T_1}$ on the complexity of \textbf{Algorithm 1} and the effect of ${T_2}$ on the complexity of the modified \textbf{Algorithm 1} are little. The complexities of  \textbf{Algorithm 1} and the modified \textbf{Algorithm 1} largely lie in the complexity of iteration and the per-iteration computation cost. It is also observed that the complexity of \textbf{Algorithm 1} under the bounded CSI error model is lower than that of the modified \textbf{Algorithm 1} under the probabilistic CSI error model.
\section{Simulation Results}
In this section, simulation results are presented to illustrate the performance of the proposed robust secure beamforming schemes. Performance comparisons are made for robust secure beamforming schemes under the bounded CSI error model and robust secure beamforming schemes under the probabilistic CSI error model. The simulation results are based on the simulation settings given in Table II.

\begin{table}
\begin{center}
\caption{Simulation Parameters}
\begin{tabular}{|c|c|c|}\hline
Parameters         &Values\\\hline
Number of PU receivers& $M=2$ \\\hline
Number of EHRs& $N=3$\\\hline
Energy conservation efficiency&$\eta=1$ \\\hline
Maximum tolerable interference power&$P_{In,i}=-10$ dB, $i\in\cal I$\\\hline
Maximum transmit power&$P_{th}=2$ dB\\\hline
Variances of noise&$\sigma _s^2=0.1$, $\sigma _e^2=0.1$, $\sigma _{s,p}^2=0.01$\\\hline
Minimum EHs of the SU and the EHRs&${\psi _s}=22$ dBm, $\psi _{e,k}=23$ dBm\\\hline
Channel power gains&$ {\mathbf{h}} \sim {\cal C}{\cal N}\left( {0,{\mathbf{I}}} \right)$, $\overline {{\mathbf{g}_k}} \sim {\cal C}{\cal N}\left( {0,{\mathbf{I}}} \right), \ k \in \cal K$, $\overline {{\mathbf{q}_i}} \sim {\cal C}{\cal N}\left( {0,{0.1\mathbf{I}}} \right), \ i \in \cal I$\\\hline
Maximum outage probabilities&$\varpi _r=\varpi _{e,k}=\varpi _{I,i}=0.05, k \in \cal {K},\ i \in \cal I$\\\hline
\end{tabular}
\end{center}
\end{table}

The covariance matrices of $\Delta {\mathbf{g}_k}, k\in \cal K$ and $\Delta {\mathbf{q}_i}, i\in\cal I$ under the probabilistic CSI error model are ${\varepsilon _{e,k}^2\mathbf{I}}$ and ${\varepsilon _{q,i}^2\mathbf{I}}$, where $\varepsilon _{e,k}^2$ and $\varepsilon _{q,i}^2$ are the variance of the corresponding CSI error, respectively. According to \cite{Q. Li2}-\cite{Z. Chu2}, in order to provide a fair comparison between the performance under the bounded CSI model with that under the probabilistic CSI error model, the radiuses of the uncertainty regions, ${\mathbf{\Psi} _{e,k}} $ and ${\mathbf{\Psi} _{P,i}}$ are set based on
\begin{subequations}
\begin{align}\label{27}\
&{\xi _{e,k}} = \sqrt {\frac{{\varepsilon _{e,k}^2F_{2{N_t}}^{ - 1}\left( {1 - {\varpi _{e,k}}} \right)}}{2}} \\
&{\xi _{P,i}} = \sqrt {\frac{{\varepsilon _{q,i}^2F_{2{N_t}}^{ - 1}\left( {1 - {\varpi _{q,i}}} \right)}}{2}}
\end{align}
\end{subequations}
where $F_{2{N_t}}^{ - 1}\left( {\cdot} \right)$ denotes the inverse cumulative distribution function (CDF) of the (central) Chi-square distribution with $2N_t$ degrees of freedom.

\begin{figure}[htb]
\centering
\includegraphics[width=3.2 in,height=2.8 in]{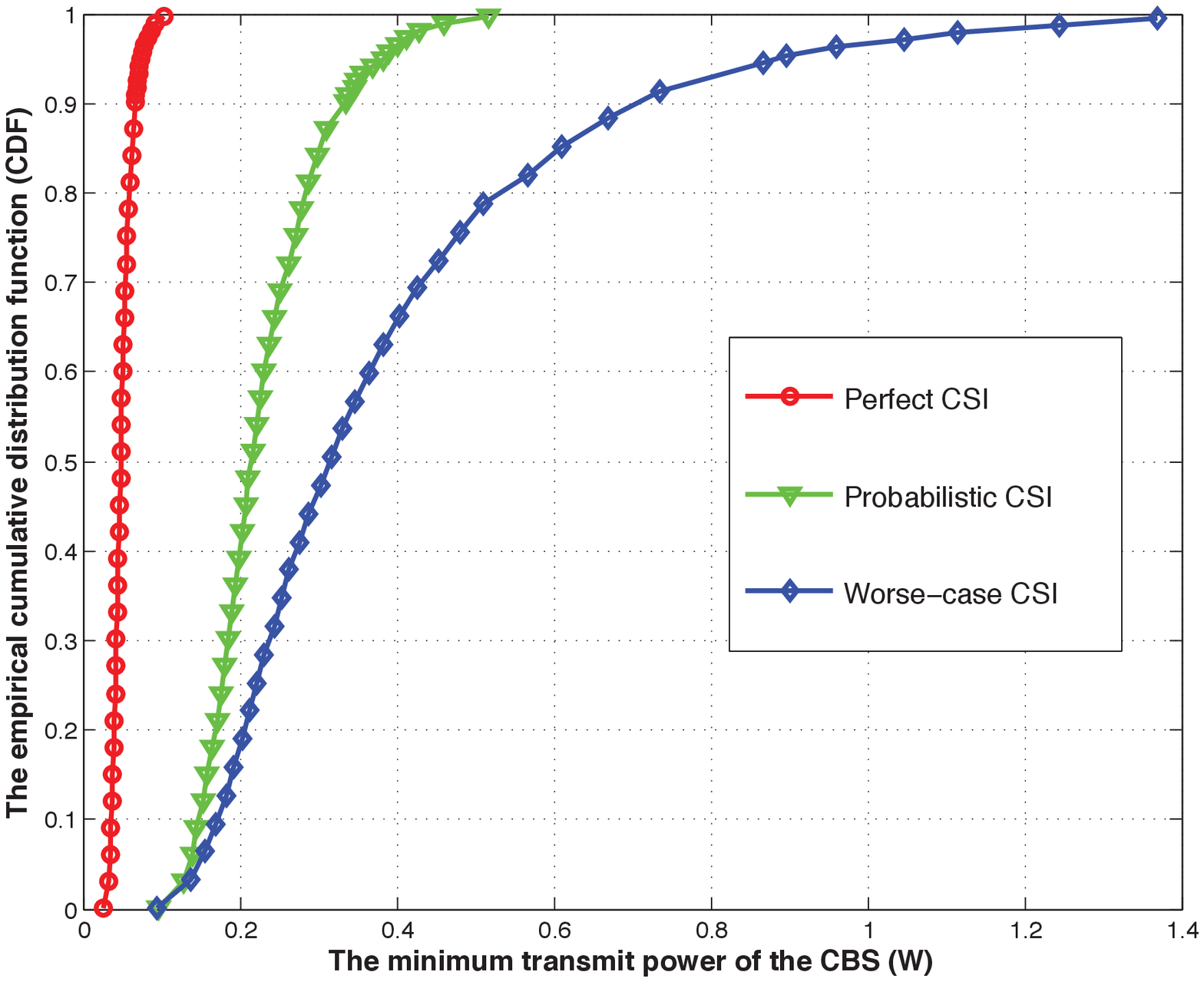}
\includegraphics[width=3.2 in,height=2.8 in]{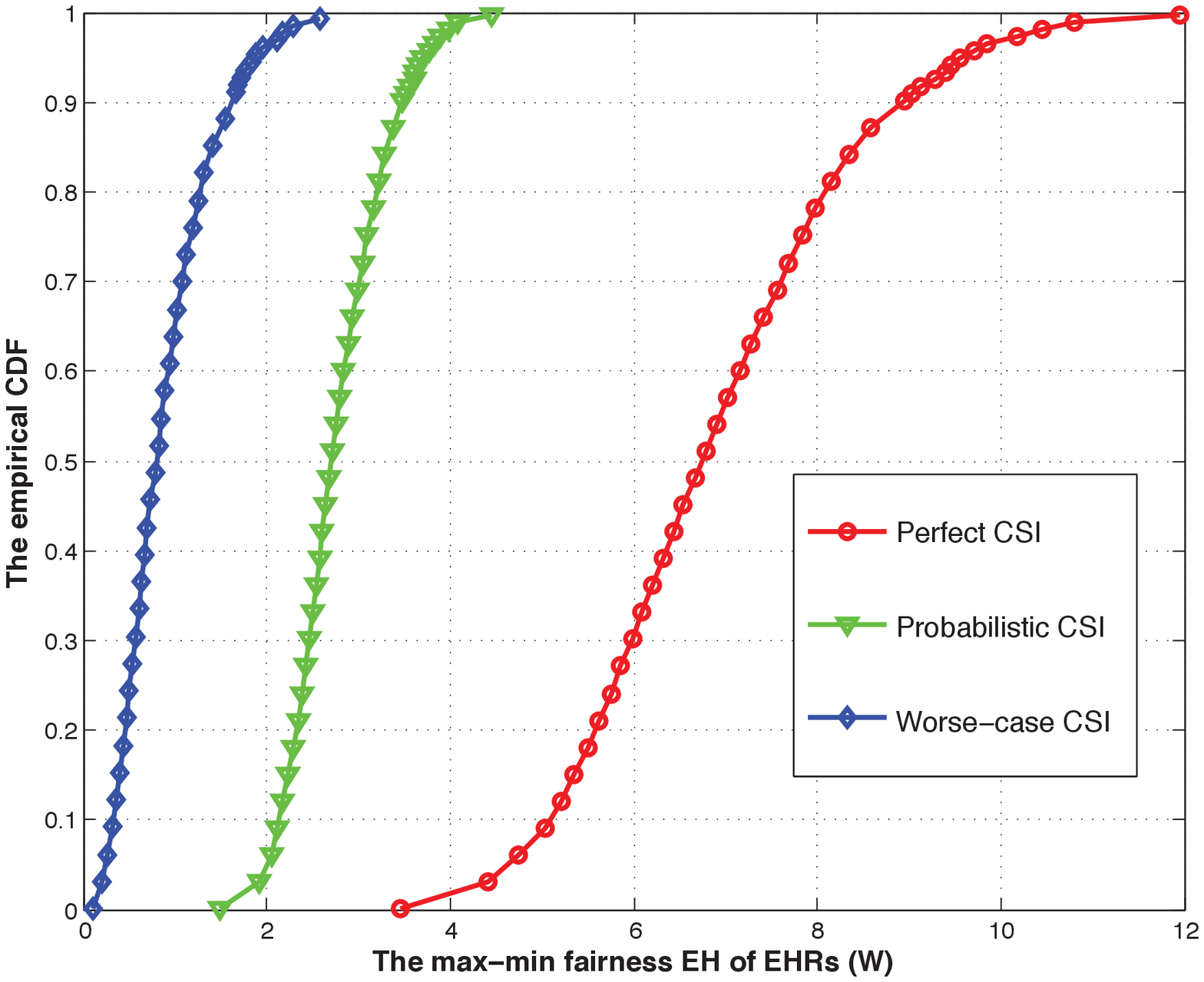}
\put(-350,-10){\footnotesize{(a)}}
\put(-115,-10){\footnotesize{(b)}}
\caption{ (a) The empirical CDF of the minimum transmit power of the CBS under different CSI scenarios, $R_{\min}=1.5$ bits/s/Hz; (b) The empirical CDF of the max-min fairness energy harvested by EHRs under different CSI scenarios, $R_{\min}=1.5$ bits/s/Hz.} \label{fig1}
\end{figure}
Figure 2 shows the empirical CDF of the minimum transmit power of the CBS and the empirical CDF of the max-min fairness energy harvested by EHRs under different CSI error scenarios. The minimum secrecy rate is set as $R_{\min}=1.5$ bits/s/Hz. $\varepsilon _{e,k}^2$ for all $k$ and $\varepsilon _{q,i}^2$ for all $i$ are set as $\varepsilon _{e,k}^2=0.001$ and $\varepsilon _{q,i}^2=0.0001$, respectively. The number of the transmit antennas of the CBS is $N_t=10$. The empirical CDFs shown in Fig. 2 are obtained by using 1,000 channel realizations. It is seen from Fig. 2(a) that the required transmit power under the perfect CSI is the least among all CSI scenarios. The reason is that the CBS under the perfect CSI scenario does not require additional transmit power to overcome channel uncertainty. As shown in Fig. 2(b), the max-min fairness energy of EHRs obtained under the perfect CSI is the largest compared with those attained under other scenarios. This is due to the fact that the CBS requires more transmit power to guarantee that the SU receiver obtains the minimum secrecy rate when the CSI is imperfect. It is interesting to observe from Fig. 2(a) that the transmit power required under the probabilistic CSI error model is less than that under the bounded CSI error model, and from Fig. 2(b) that the max-min fairness energy of EHRs attained under the probabilistic CSI error model is larger than that obtained under the bounded CSI error model. This can be explained by the fact that the constraints associated with the secrecy rate, the EH of EHRs and the interference power under the probabilistic CSI error model is looser than those under the bounded CSI error model.

\begin{figure}[!t]
\centering
\includegraphics[width=2.8 in,height=2.5 in]{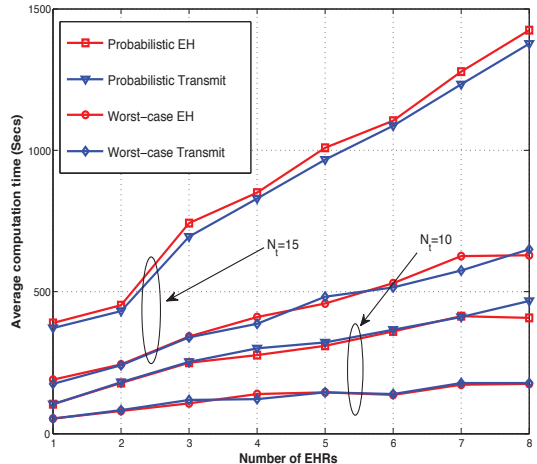}
\caption{Average computation time of different optimization problems under different CSI error models, $R_{\min}=1.5$ bits/s/Hz, $N_t=10$ or $N_t=15$.} \label{fig.1}
\end{figure}
A comparison is presented in Fig. 3 for the average computation times of different optimization problems (i.e., the transmit power minimization problem and the max-min EH fairness problem) under the probabilistic CSI error model with those under the bounded CSI error model. The results are obtained by using a computer with 32-bit Intel(R) Core(TM) i5-3230M, $4$ GB RAM. It is seen that the average computation time of the transmit power minimization problem is almost the same as that of the max-min EH fairness problem under the probabilistic CSI error model. This phenomenon can also be seen under the bounded CSI error model. The reason is that the complexities of different optimization problems presented in Section III and Section IV under the same CSI error model are almost the same. As shown in Fig. 3, the average computation time of the optimization problem under the probabilistic CSI error model is higher than that of the corresponding optimization problem under the bounded error model. This indicates that the complexity of the optimization problem under the probabilistic CSI error model is higher than that of the corresponding optimization problem under the bounded error model. Moreover, it is observed that the average computation time greatly increases with the number of the transmit antennas of the CBS irrespective of the optimization problems and the CSI error models. This is in complete agreement with our analysis presented in Section IV.

\begin{figure}[!t]
\centering
\includegraphics[width=2.8 in,height=2.5 in]{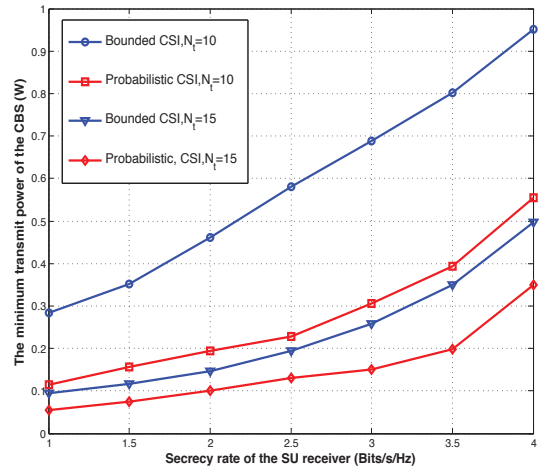}
\caption{The minimum transmit power of the CBS versus the secrecy rate of the SU receiver under different CSI error models, $N_t=10$ or $N_t=15$.} \label{fig.1}
\end{figure}
Figure 4 shows the minimum transmit power of the CBS versus the secrecy rate of the SU receiver under different CSI error models. $N_t$ is set as $N_t=10$ or $N_t=15$. It is seen that the minimum transmit power of the CBS increases with the minimum secrecy rate requirement, regardless of the CSI error model. The reason is that a higher transmit power is required to guarantee the increasing secrecy rate requirement. It is also observed that the transmit power of the CBS decreases with the increase of the number of the transmit antennas of the CBS. It is explained by the fact that an increase of the number of antennas increases degrees of freedom of the transmit power allocation. Finally, it is also seen that the minimum transmit power of the CBS under the bounded CSI error model is larger than that under the probabilistic CSI error model.

\begin{figure}[!t]
\centering
\includegraphics[width=2.8 in,height=2.5 in]{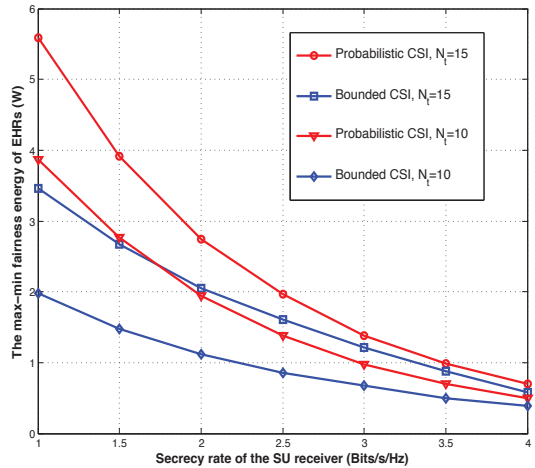}
\caption{The max-min fairness energy harvested by EHRs versus the secrecy rate of the SU receiver under different CSI error models, $N_t=10$ or $N_t=15$.} \label{fig.1}
\end{figure}
Figure 5 compares the max-min fairness energy harvested by EHRs achieved under the probabilistic CSI error model with that obtained under the bounded CSI error model. The number of the transmit antennas of the CBS is set as $N_t=10$ or $N_t=15$. It is seen that the max-min fairness energy of EHRs decreases with the increase of the minimum secrecy rate requirement of the SU receiver irrespective of the CSI error models. This indicates that there is a tradeoff between the max-min fairness energy harvested by EHRs and the secrecy rate of the SU receiver. It is interesting to observe that the gap between the max-min fairness energy of EHRs achieved under the probabilistic CSI error model and that obtained under the bounded CSI model decreases with the increase of the minimum secrecy rate requirement of the SU receiver. This phenomenon is due to the transmit power constraint, which limits the maximum transmit power of the CBS. As shown in Fig. 5, the max-min fairness energy harvested by EHRs increases with the number of the transmit antennas of the CBS. This can be explained by the fact that the diversity gain increases when the number of the transmit antennas of the CBS is increased. Finally, it is also observed that the max-min fairness energy of EHRs achieved under the probabilistic CSI error model is larger than that achieved under the bounded CSI error model.

\begin{figure}[htb]
\centering
\includegraphics[width=3.2 in,height=2.8 in]{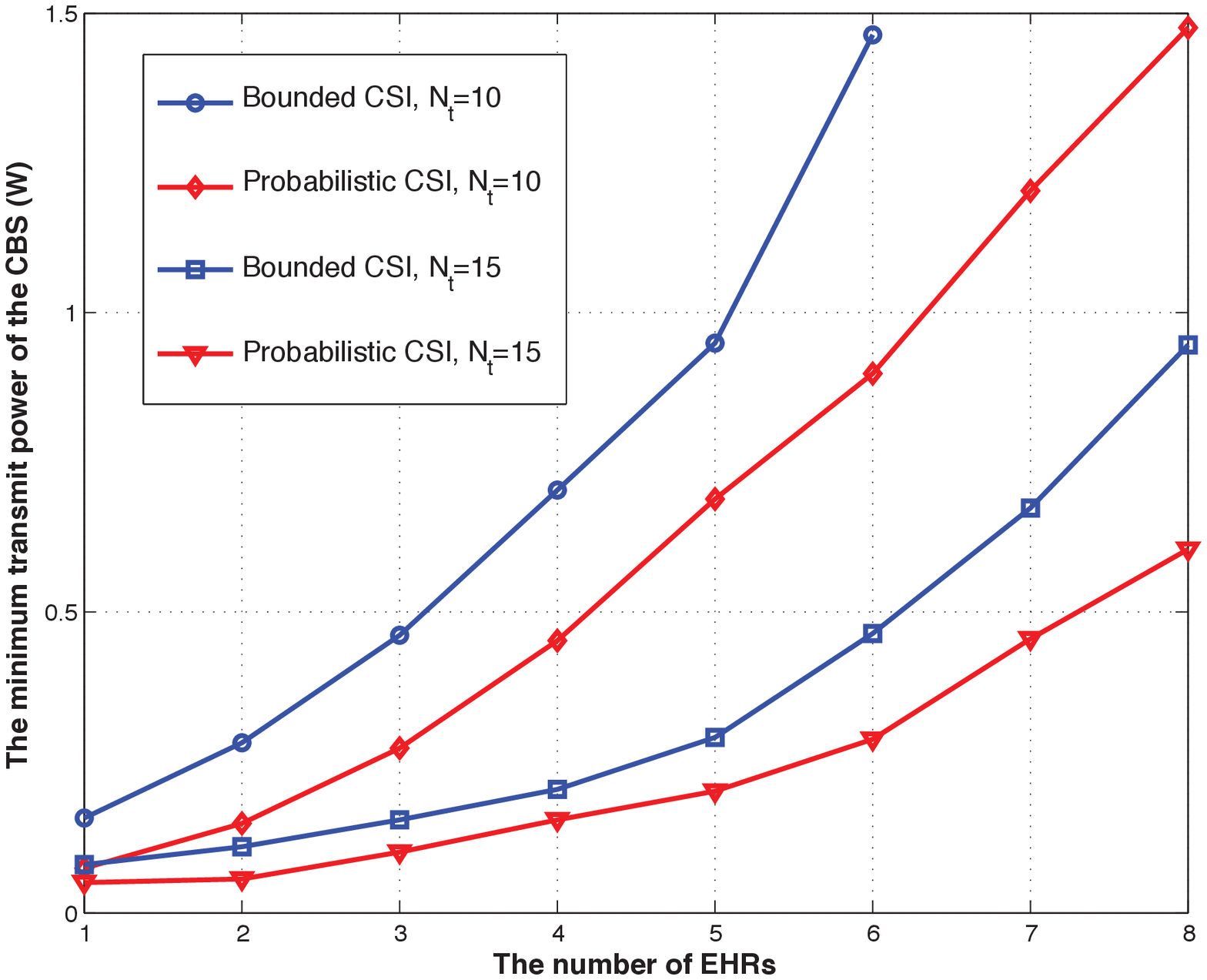}
\includegraphics[width=3.2 in,height=2.8 in]{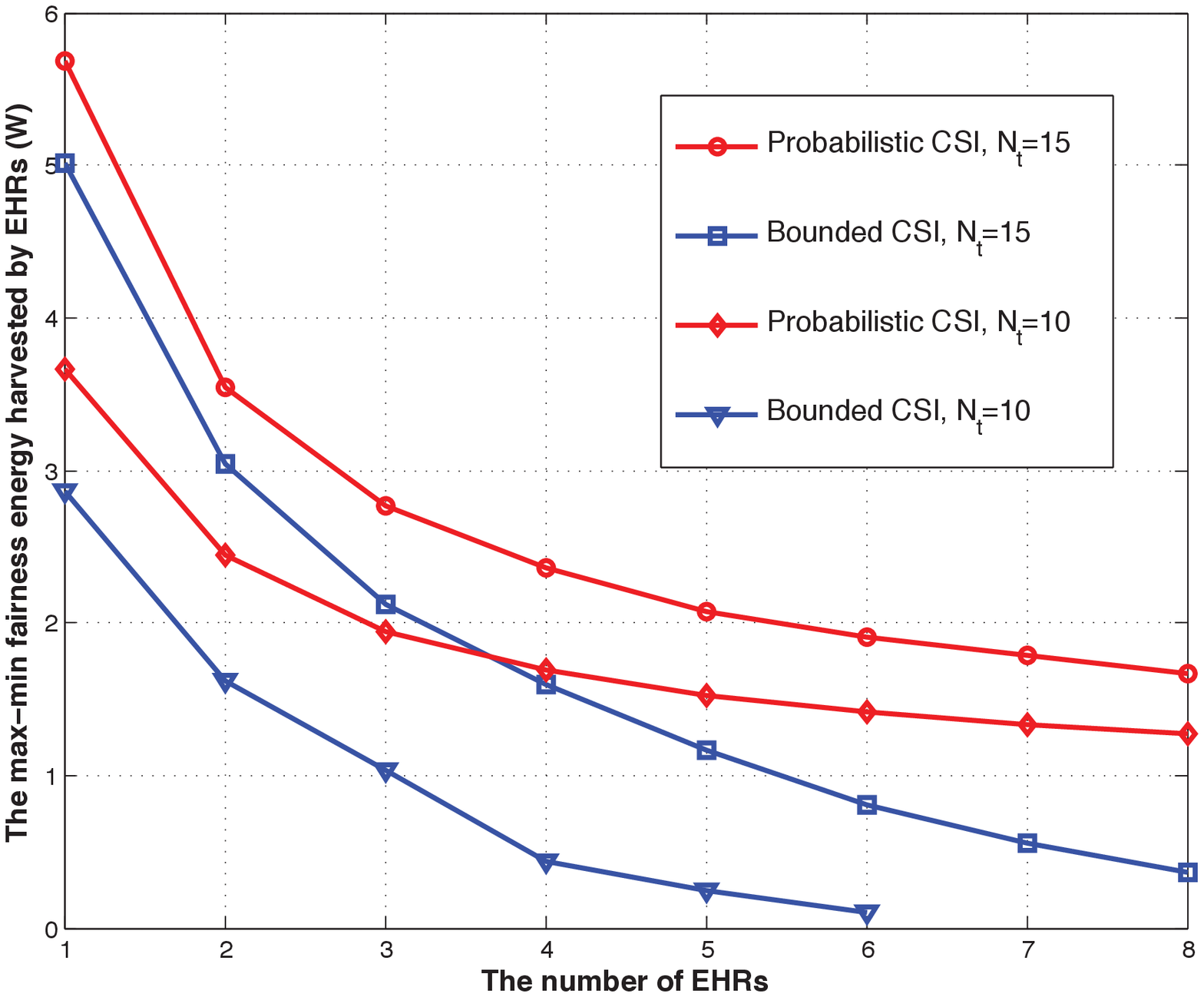}
\put(-350,-10){\footnotesize{(a)}}
\put(-115,-10){\footnotesize{(b)}}
\caption{ (a) The minimum transmit power of the CBS versus the number of EHRs under different CSI error scenarios, $R_{\min}=2$ bits/s/Hz, $N_t=10$ or $N_t=15$; (b) The max-min fairness energy harvested by EHRs versus the number of EHRs under different CSI error scenarios, $R_{\min}=2$ bits/s/Hz, $N_t=10$ or $N_t=15$.} \label{fig1}
\end{figure}
Figure 6 presents the minimum transmit power of the CBS and the max-min fairness energy harvested by EHRs versus the number of EHRs under different CSI error models. The minimum secrecy rate of the SU receiver is set as $R_{\min}=2$ bits/s/Hz. The number of the transmit antennas of the CBS is $N_t=10$ or $N_t=15$, respectively. It is seen from Fig. 6(a) that the minimum transmit power of the CBS increases with the number of EHRs irrespective of the CSI error models. The reason is that the transmit power of the CBS with respect to the transmit power minimization problem is required to be increased in order to satisfy the minimum secrecy rate requirement of the SU receiver when the number of the EHRs increases. It is observed from Fig. 6(b) that the max-min fairness energy harvested by EHRs decreases with the increase of the number of EHRs. This can be explained by the fact that the robust secure beamforming is required to be designed to guarantee the secrecy rate of the SU receiver, which is susceptible to the number of the EHRs. As shown in Fig. 6(a) and Fig. 6(b), when the number of the transmit antennas of the CBS is $N_t=10$, the transmit power minimization problem and the max-min fairness EH problem under the bounded CSI error model are infeasible when the number of EHRs is larger than $6$. The reason is that the minimum secrecy rate of the SU receiver cannot be satisfied due to the maximum transmit power constraint. This is consistent with our analysis presented in Section III. It is also seen from Fig. 6 that a performance gain can be obtained under the probabilistic error model compared with that attained under the bounded error model irrespective of the number of EHRs. Specifically, the minimum transmit power of the CBS required under the probabilistic error model is smaller than that required under the bounded model, and the max-min energy harvested by EHRs under the probabilistic error model is larger than that obtained under the bounded model irrespective of the number of EHRs.
\section{Conclusion}
Secure communication was studied for a MISO CR downlink network with SWIPT. Robust secure beamforming and power splitting ratio were jointly designed under the imperfect CSI in order to guarantee secure communication and energy harvested by EHRs. The transmit power of the CBS and the max-min fairness energy harvested by EHRs were optimized under the bounded CSI error model and the probabilistic CSI error model. A one-dimensional search algorithm was proposed to solve these challenging non-convex problems. The inner component of the proposed algorithm is only required to solve the convex problems. It was proved that the optimal robust secure beamforming vector and the optimal AN covariance matrix can always be found under the bounded CSI error model. The optimal robust secure beamforming vector under the probabilistic CSI error model cannot be guaranteed and a suboptimal beamforming scheme was found. It was shown that a performance gain can be achieved under the probabilistic CSI error model compared with that under the bounded CSI error model at the cost of the implementation complexity. A tradeoff was found between the secure rate of the SU receiver and the energy harvested by EHRs under a max-min fairness criterion.
\appendices

\section{Proof of Constraint $C1$}

By introducing a slack variable $\beta$, one can rewrite $C1$ as
\begin{subequations}
\begin{align}\label{27}\
&{\log _2}\left( {1 + \frac{{\rho {\mathbf{h}^\dag }\mathbf{W}\mathbf{h}}}{{\rho \left( {{\mathbf{h}^\dag }\mathbf{\Sigma} \mathbf{h} + \sigma _s^2} \right) + \sigma _{s,p}^2}}} \right) - {\log _2}\beta  \ge {R_{\min }}\\
&1 + \frac{{\mathbf{g}_k^\dag \mathbf{W}{\mathbf{g}_k}}}{{\mathbf{g}_k^\dag \mathbf{\Sigma} {\mathbf{g}_k} + \sigma _e^2}} \le \beta ,\ \forall \Delta {\mathbf{g}_k} \in {\mathbf{\Psi} _{e,k}},\ k \in K.
\end{align}
\end{subequations}
Let $t=1/\rho$, where $t>1$. Eq. $\left(10\rm{a}\right)$ is obtained by performing some algebraic manipulations on $\left(36\rm{a}\right)$. Applying \textbf{Lemma 1}, using the fact that $\Delta \mathbf{g}_k^\dag \Delta {\mathbf{g}_k} \le \xi _{e,k}^2,k \in \cal K$, and substituting ${g_k} = \overline {{g_k}}  + \Delta {g_k}$ into $\left(36\rm{b}\right)$, one has
\begin{align}\label{27}\ \notag
& \Delta \mathbf{g}_k^\dag \left( {\mathbf{W} - \left( {\beta  - 1} \right) \mathbf{\Sigma} } \right)\Delta {\mathbf{g}_k} + 2{\mathop{\rm Re}\nolimits} \left( {{{\overline {{\mathbf{g}_k}} }^\dag }\left( {\mathbf{W} - \left( {\beta  - 1} \right) \mathbf{\Sigma} } \right)\Delta {\mathbf{g}_k}} \right) \\
&\ \ \ \ \ \ \ \ \ \ \ \ \ \ \  \ \ \ \ \ \ \ \ \ \ \ \ \ \ \ \ \ \  + {\overline {{\mathbf{g}_k}} ^\dag }\left( {\mathbf{W} - \left( {\beta  - 1} \right) \mathbf{\Sigma} } \right)\overline {{\mathbf{g}_k}}  - \left( {\beta  - 1} \right)\sigma _e^2 \le 0
\end{align}
and $\left(10\rm{b}\right)$ holds  when ${\omega _k} \ge 0$. This completes the proof.
\section{Proof of Theorem 1}
The proof for Theorem 1 is based on the KKT optimality conditions of $\textbf{P}_{{2}}$. Let $\mathbf{\Xi}$ denote a collection of all the dual and primal variables related to $\textbf{P}_{{2}}$. Then, the Lagrangian of $\textbf{P}_{{2}}$ is given by
\begin{align}\label{27}\ \notag
\cal {L} \left(  \mathbf{\Xi} \right) =& \text{Tr}\left( {\mathbf{W} + \mathbf{\Sigma} } \right) - \alpha \left\{ {\text{Tr}\left\{ {\left( {\mathbf{W} + \left( {1 - {2^{{R_{\min }}}}\beta } \right)\mathbf{\Sigma} } \right)\mathbf{H}} \right\} + \left( {1 - {2^{{R_{\min }}}}\beta } \right)\left( {\sigma _s^2 + t\sigma _{s,p}^2} \right)} \right\}\\ \notag
& - \sum\limits_{k = 1}^K {\text{Tr}\left\{ {{\mathbf{A}_k}{\mathbf{\Gamma} _k}\left( {{\omega _k},\mathbf{W},\mathbf{\Sigma} ,\beta } \right)} \right\}}  - \sum\limits_{k = 1}^K {\text{Tr}\left\{ {{\mathbf{B}_k}{\mathbf{\Gamma} _k}\left( {{\mu _k},\mathbf{W},\mathbf{\Sigma} } \right)} \right\}}  - \sum\limits_{i = 1}^M {\text{Tr}\left\{ {{\mathbf{D}_i}{\mathbf{\Gamma} _i}\left( {{\delta _i},\mathbf{W},\mathbf{\Sigma} } \right)} \right\}} \\ \notag
& - {\nu _1}\left\{ {\left( {\text{Tr}\left( {\mathbf{W}\mathbf{H} + \mathbf{\Sigma}\mathbf{H}} \right) + \sigma _s^2} \right) - \left( {1 + \frac{1}{{t - 1}}} \right)\frac{1}{\eta }{\psi _s}} \right\} + {\nu _2}\left( {\text{Tr}\left( {\mathbf{W} + \mathbf{\Sigma} } \right) - {P_{th}}} \right) \\
 &- \text{Tr}\left( {\mathbf{W}\mathbf{Y}} \right) - \text{Tr}\left( {\mathbf{\Sigma} \mathbf{Z}} \right)+ \Lambda
\end{align}
where $ \alpha\in \mathbb{R}_{+}$, $ \nu _1\in \mathbb{R}_{+}$ and $ \nu _2\in \mathbb{R}_{+}$, are the dual variables with respect to $\left(10\rm{a}\right)$, $\left(12\rm{b}\right)$ and $C5$, respectively. $\Lambda$ denotes the collection of terms involving the variables, which are not related to the proof. ${\mathbf{A}_k}\in \mathbb{H}_+^{N}$, ${\mathbf{B}_k}\in \mathbb{H}_+^{N}$, ${\mathbf{D}_i}\in \mathbb{H}_+^{N}$, $\mathbf{Y}\in \mathbb{H}_+^{N}$ and $\mathbf{Z}\in \mathbb{H}_+^{N}$ are the dual variables with respect to $\left(10\rm{b}\right)$, $\left(11\rm{a}\right)$, $\left(11\rm{b}\right)$ and $\left(8\rm{i}\right)$, respectively. Let ${ \mathbf{\Lambda} _k} = [ {\begin{array}{*{20}{c}}
 \mathbf{I}&{\overline {{ \mathbf{g}_k}} }
\end{array}} ]
$ and ${\mathbf{\Upsilon} _i} = [ {\begin{array}{*{20}{c}}
\mathbf{I}&{\overline {{\mathbf{q}_i}} }
\end{array}} ]
$. Then, ${\mathbf{\Gamma} _k}\left( {{\omega _k},\mathbf{W},\mathbf{\Sigma} ,\beta } \right)$, ${\mathbf{\Gamma} _k}\left( {{\mu _k},\mathbf{W},\mathbf{\Sigma} } \right)$ and ${\mathbf{\Gamma} _i}\left( {{\delta _i},\mathbf{W},\mathbf{\Sigma} } \right)$ can be rewritten as
\begin{subequations}
\begin{align}\label{27}\
&{\mathbf{\Gamma} _k}\left( {{\omega _k},\mathbf{W},\mathbf{\Sigma} ,\beta } \right) = \left[ {\begin{array}{*{20}{c}}
{{\omega _k}\mathbf{I}}&0\\
0&{\left( {\beta  - 1} \right)\sigma _e^2 - {\omega _k}\xi _{e,k}^2}
\end{array}} \right] - \mathbf{\Lambda} _k^H\left( {\mathbf{W}- \left( {\beta  - 1} \right)\mathbf{\Sigma} } \right){\mathbf{\Lambda} _k}
\\
& {\mathbf{\Gamma} _k}\left( {{\mu _k},\mathbf{W},\mathbf{\Sigma} } \right)= \left[ {\begin{array}{*{20}{c}}
{{\mu _k}\mathbf{I}}&0\\
0&{\sigma _e^2 - {\psi _{e,k}}{\eta ^{ - 1}} - {\mu _k}\xi _{e,k}^2}
\end{array}} \right] + \mathbf{\Lambda} _k^H\left( {\mathbf{W} + \mathbf{\Sigma} } \right){\mathbf{\Lambda} _k}, k\in \cal K\\
&{\mathbf{\Gamma} _i}\left( {{\delta _i},\mathbf{W},\mathbf{\Sigma} } \right) = \left[ {\begin{array}{*{20}{c}}
{{\delta _i}\mathbf{I}}&0\\
0&{{P_{In,i}} - {\delta _i}\xi _{P,i}^2}
\end{array}} \right] - \mathbf{\Upsilon} _i^H\left( {\mathbf{W} + \mathbf{\Sigma} } \right){\mathbf{\Upsilon} _i}, i\in \cal I.
\end{align}
\end{subequations}
The partial KKT conditions related to the proof can be given as
\begin{subequations}
\begin{align}\label{27}\
&\mathbf{I} - \alpha \mathbf{H} + \sum\limits_{k = 1}^K {{\mathbf{\Lambda} _k}\left( {{\mathbf{A}_k} - {\mathbf{B}_k}} \right)\mathbf{\Lambda}_k^H}  + \sum\limits_{i = 1}^M {\mathbf{\Upsilon} _i{\mathbf{D}_i}\mathbf{\Upsilon} _i^H}  - {\nu _1}\mathbf{H} + {\nu _2}\mathbf{I} - \mathbf{Y} = \mathbf{0}
\\
& \mathbf{Y}\mathbf{W} = \mathbf{0}\\
&t = \frac{{\sqrt {\eta \alpha \left( { {2^{{R_{\min }}}}\beta } -1\right)} {\sigma _{s,p}} + \sqrt {{\nu _1}{\psi _s}} }}{{\sqrt {\eta \alpha \left( { {2^{{R_{\min }}}}\beta }-1 \right)} {\sigma _{s,p}}}}\\
&{\mathbf{A}_k},{\mathbf{B}_k},{\mathbf{D}_i}\succeq\mathbf{0}, \alpha, {\nu _1},{\nu _2} \ge 0.
\end{align}
\end{subequations}
Since $t>1$, it can be obtained from $\left(40\rm{c}\right)$ that $\alpha>0$ and $\nu _1>0$. Right-multiplying $\left(40\rm{a}\right)$ by $\mathbf{W}$ and combining $\left(40\rm{b}\right)$, one has
\begin{align}\label{27}\
\left( {\left( {1 + {\nu _2}} \right)\mathbf{I} + \sum\limits_{i = 1}^M {{\mathbf{\Upsilon} _i}{\mathbf{D}_i}\mathbf{\Upsilon} _i^H}  + \sum\limits_{k = 1}^K {{\mathbf{\Lambda}_k}\left( {{\mathbf{A}_k} - {\mathbf{B}_k}} \right)\mathbf{\Lambda} _k^H} } \right)\mathbf{W} = \left( {\alpha  + {\nu _1}} \right)\mathbf{H}\mathbf{W}.
\end{align}
According to $\left(41\right)$, one has
\begin{subequations}
\begin{align}\label{27}\
&\text{Rank}\left\{\left( {\left( {1 + {\nu _2}} \right)\mathbf{I} + \sum\limits_{i = 1}^M {{\mathbf{\Upsilon} _i}{\mathbf{D}_i}\mathbf{\Upsilon} _i^H}  + \sum\limits_{k = 1}^K {{\mathbf{\Lambda}_k}\left( {{\mathbf{A}_k} - {\mathbf{B}_k}} \right)\mathbf{\Lambda} _k^H} } \right)\mathbf{W} \right\}= \text{Rank}\left\{\left( {\alpha  + {\nu _1}} \right)\mathbf{H}\mathbf{W}\right\}
\\
& \text{Rank}\left\{\left( {\alpha  + {\nu _1}} \right)\mathbf{H}\mathbf{W}\right\}\leq1.
\end{align}
\end{subequations}
According to $\left(40\rm{a}\right)$, one has
\begin{align}\label{27}\
\left( {\left( {1 + {\nu _2}} \right)\mathbf{I} + \sum\limits_{i = 1}^M {{\mathbf{\Upsilon} _i}{\mathbf{D}_i}\mathbf{\Upsilon} _i^H}  + \sum\limits_{k = 1}^K {{\mathbf{\Lambda}_k}\left( {{\mathbf{A}_k} - {\mathbf{B}_k}} \right)\mathbf{\Lambda} _k^H} } \right) = \mathbf{Y}+\left( {\alpha  + {\nu _1}} \right)\mathbf{H}.
\end{align}
Since $\mathbf{Y}+\left( {\alpha  + {\nu _1}} \right)\mathbf{H}\succ \mathbf{0}$, one obtains the following relationship
\begin{align}\label{27}\ \notag
\text{Rank}\left(\mathbf{W}\right)&=\text{Rank}\left\{\left( {\left( {1 + {\nu _2}} \right)\mathbf{I} + \sum\limits_{i = 1}^M {{\mathbf{\Upsilon} _i}{\mathbf{D}_i}\mathbf{\Upsilon} _i^H}  + \sum\limits_{k = 1}^K {{\mathbf{\Lambda}_k}\left( {{\mathbf{A}_k} - {\mathbf{B}_k}} \right)\mathbf{\Lambda} _k^H} } \right)\mathbf{W} \right\}\\
&= \text{Rank}\left\{\left( {\alpha  + {\nu _1}} \right)\mathbf{H}\mathbf{W}\right\}\leq1.
\end{align}
Thus, if $\textbf{P}_{{2}}$ is feasible and $R_{\min}>0$, the rank of $\mathbf{W}$ is one. The proof is completed.


\begin{thebibliography}{20}
\bibitem{Haykin}
S. Haykin, \lq\lq Cognitive radio: brain-empowered wireless communications,\rq\rq \  \emph{IEEE J. Sel. Areas Commun.}, vol. 23, no. 2, pp. 201-220, Feb. 2005.
\bibitem{O. Akan}
O. Akan, O. Karli, and O. Ergul, \lq\lq Cognitive radio sensor networks,\rq\rq \  \emph{IEEE Netw.}, vol. 23, no. 4, pp. 34-40, Jul. 2009.
\bibitem{X. Huang}
X. Huang, T. Han, and N. Ansari, \lq\lq On green energy powered cognitive radio networks,\rq\rq \  \emph{IEEE Commun. Surveys Tuts.}, vol. 17, no. 2, pp. 827-842, Second Quarter, 2015.
\bibitem{S. Chen}
S. Chen and J. Zhao, \lq\lq The requirements, challenges and technologies for 5G of terrestrial mobile telecommunication,\rq\rq \  \emph{IEEE Commun. Mag.}, vol.52, no. 5, pp. 36-43, May 2014.
\bibitem{Y. Chen}
Y. Chen, S. Zhang, S. Xu, and G. Li, \lq\lq Fundamental trade-offs on green wireless networks,\rq\rq \ \emph{IEEE Commun. Mag.}, vol. 49, no. 6, pp. 30-37, Jun. 2011.
\bibitem{C. Jiang}
C. Jiang, H. Zhang, Y. Ren, and H. Chen, \lq\lq Energy-efficient non-cooperative cognitive radio networks: micro, meso, and macro views,\rq\rq \ \emph{IEEE Commun. Mag.}, vol. 52, no. 7, pp. 14-20, Jun. 2014.
\bibitem{Y. Pei11}
Y. Pei, Y. C. Liang, and K. Teh, \lq\lq Energy-efficient design of sequential channel sensing in cognitive radio networks: optimal sensing strategy, sensing order,\rq\rq \ \emph{IEEE J. Sel. Areas Commun.}, vol. 29, no. 8, pp. 1648-1659, Sep. 2011.
\bibitem{C. Luo}
C. Luo, G. Min, F. Yu, M. Chen, L. L. Yang, and V. C. M. Leung, \lq\lq Energy-efficient distributed relay and power control in cognitive radio cooperative communications,\rq\rq \ \emph{IEEE J. Sel. Areas Commun.}, vol. 31, no. 11, pp. 2442-2452, Nov. 2013.
\bibitem{F. H. Zhou}
F. Zhou, N. C. Beaulieu, Z. Li, J. Si, and P. Qi, \lq\lq Energy-efficient optimal power allocation for fading cognitive radio channels: ergodic capacity, outage capacity and minimum-rate capacity,\rq\rq \ \emph{IEEE Trans. Wireless Commun.}, to be published, 2015.
\bibitem{A. Alabbasi}
A. Alabbasi, Z. Rezki, and B. Shihada, \lq\lq Energy efficient resource allocation for cognitive radios: a generalized sensing analysis,\rq\rq \ \emph{IEEE Trans. Wireless Commun.}, vol. 14, no. 5, pp. 2455-2469, May 2015.
\bibitem{Y. Dong}
Y. Dong, M. J. Hossain, and J. Cheng, \lq\lq Joint power control and time switching for SWIPT systems with heterogeneous QoS requirements,\rq\rq \ \emph{IEEE Commun. Lett.}, to be published, 2015.
\bibitem{I. Krikidis}
I. Krikidis, S. Timotheou, S. Nikolaou, G. Zheng, D. W. K. Ng, and R. Schober, \lq\lq Simultaneous wireless information and power transfer in modern communication systems,\rq\rq \ \emph{IEEE Commun. Mag.}, vol. 52, no. 11, pp. 104-110, Nov. 2014.
\bibitem{X. Lu}
X. Lu, P. Wang, D. Niyato, D. I. Kim, and Z. Han, \lq\lq Wireless networks with RF energy harvesting: A contemporary survey,\rq\rq \ \emph{IEEE Commun. Surveys Tuts.}, vol. 17, pp. 757-789, Second Quarter, 2015.
\bibitem{D. W. K. Ng1}
D. W. K. Ng, E. S. Lo, and R. Schober, \lq\lq Multi-objective resource allocation for secure communication in cognitive radio networks with wireless information and power transfer,\rq\rq \ \emph{IEEE Trans. Veh. Technol.}, to be published, 2015.
\bibitem{B. Fang}
B. Fang, Z. Qian, W. Zhang, and W. Shao, \lq\lq An-aided secrecy precoding for SWIPT in cognitive MIMO broadcast channels,\rq\rq \ \emph{IEEE Commun. Lett.}, vol. 19, no. 9, pp. 1632-1635, Sep. 2015.
\bibitem{R. Liu}
R. Liu and W. Trappe, \emph{Securing Wireless Communications at the Physical Layer}, 1st ed. New York: Springer-Verlag, 2009.
\bibitem{Y. Pei}
Y. Pei, Y. C. Liang, K. C. Teh, and K. H. Li, \lq\lq Secure communication over MISO cognitive radio channels,\rq\rq \ \emph{IEEE Trans. Wireless Commun.}, vol. 9, no. 4, pp. 1494-1502, Apr. 2010.
\bibitem{Y. Pei1}
Y. Pei, Y. C. Liang, K. Teh, and K. H. Li, \lq\lq Secure communication in multiantenna cognitive radio networks with imperfect channel state information,\rq\rq \ \emph{IEEE Trans. Signal Process.}, vol. 59, no. 4, pp. 1683-1693, Apr. 2011.
\bibitem{L. Zhang}
L. Zhang, R. Zhang, Y. C. Liang, Y. Xin, and S. Cui, \lq\lq On the relationship between the multi-antenna secrecy communications and cognitive radio communications,\rq\rq \ \emph{IEEE Trans. Commun.}, vol. 58, no. 6, pp. 1877-1886, Jun. 2010.
\bibitem{S. Ma}
S. Ma and D. Sun, \lq\lq Chance constrained robust beamforming in cognitive radio networks,\rq\rq \ \emph{IEEE Commun. Lett.}, vol. 17, no. 1, pp. 67-70, Jan. 2013.
\bibitem{C. Wang}
C. Wang and H. M. Wang, \lq\lq On the secrecy throughput maximization for MISO cognitive radio network in slow fading channels,\rq\rq \ \emph{IEEE Trans. Inf. Forensics Security}, vol. 9, no. 11, pp. 1814-1827, Nov. 2014.
\bibitem{C. Xu}
C. Xu, Q. Zhang, Q. Li, Y. Tan, and J. Qin, \lq\lq Robust transceiver design for wireless information and power transmission in underlay MIMO cognitive radio networks,\rq\rq \ \emph{IEEE Commun. Lett.}, vol. 18, no. 9, pp. 1665-1668, Sept. 2014.
\bibitem{J. Huang}
J. Huang and L. A. Swindlehurst,\lq\lq Robust secure transmission in MISO channels based on worst-case optimization,\rq\rq \ \emph{IEEE Trans. Signal Process.}, vol. 60, no. 4, pp. 1696-1702, Apr. 2012.
\bibitem{Z. Chu}
Z. Chu, K. Cumanan, Z. Ding, M. Johnston, and S. Le Goff, \lq\lq Secrecy rate optimizations for a MIMO secrecy channel with a cooperative jammer,\rq\rq \ \emph{IEEE Trans. Vehicular Technol.}, vol. 64, no. 5, pp. 1833-1847, May 2015.
\bibitem{Q. Li1}
Q. Li and W. K. Ma, \lq\lq Optimal and robust transmit designs for MISO channel secrecy by semidefinite programming,\rq\rq \ \emph{IEEE Trans. Signal Process.}, vol. 59, no. 8, pp. 3799-3812, Aug. 2011.
\bibitem{Q. Li2}
Q. Li and W. K. Ma, \lq\lq Spatially selective artificial-noise aided transmit optimization for MISO multi-eves secrecy rate maximization,\rq\rq \ \emph{IEEE Trans. Signal Process.}, vol. 61, no. 10, pp. 2704-2717, May 2013.
\bibitem{S. Ma2}
S. Ma, M. Hong, E. Song, X. Wang, and D. Sun, \lq\lq Outage constrained robust secure transmission for MISO wiretap channels,\rq\rq \ \emph{IEEE Trans. Wireless Commun.}, vol. 13, no. 10, pp. 5558-5570, Oct. 2014.
\bibitem{K. Y. Wang}
K. Y. Wang, A. M. C. So, T. H. Chang, W. K. Ma, and C. Y. Chi, \lq\lq Outage constrained robust transmit optimization for multiuser MISO downlinks: tractable approximations by conic optimization,\rq\rq \ \emph{IEEE Trans. Signal Process.}, vol. 62, no. 21, pp. 5690-5705, Nov. 2014.
\bibitem{Z. Chu2}
Z. Chu, H. Xing, M. Johnston, and S. Le Goff, \lq\lq Secrecy rate optimizations for a MISO secrecy channel with multiple multi-antenna eavesdroppers,\rq\rq \ \emph{IEEE Trans. Wireless Commun.}, to be published, 2015.
\bibitem{Q. Li4}
Q. Li, W. K. Ma, and A. M. C. So, \lq\lq Safe convex approximation to outage-based MISO secrecy rate optimization under imperfect CSI and with artificial noise,\rq\rq \ in \emph{Proc. 45th Annual Asilomar Conference on Signals, Systems, and Computers}, Pacific Grove, CA, USA, Nov. 3-6, 2011, pp. 207-211.
\bibitem{Q. Li3}
Q. Li, W. K. Ma, and A. M. So, \lq\lq A safe approximation approch to secure outage design for MIMO wiretap channels,\rq\rq \ \emph{IEEE Signal Process. Lett.}, vol. 21, no. 1, pp. 118-121, Jan. 2014.
\bibitem{D. W. K. Ng2}
D. W. K. Ng, E. S. Lo, and R. Schober, \lq\lq Robust beamforming for secure communication in systems with wireless information and power transfer,\rq\rq \ \emph{IEEE Trans. Wireless Commun.}, vol. 13, no. 8, pp. 4599-4615, Aug. 2014.
\bibitem{D. W. K. Ng3}
D. W. K. Ng and R. Schober, \lq\lq Secure and green SWIPT in distributed antenna networks with limited backhaul capacity,\rq\rq \ \emph{IEEE Trans. Wireless Commun.}, vol. 14, no. 1, pp. 5082-5097, Sept. 2015.
\bibitem{M. R. A. Khandaker}
M. R. A. Khandaker and K. K. Wong, \lq\lq Robust secrecy beamforming with energy-harvesting eavesdroppers,\rq\rq \ \emph{IEEE Wireless Commun. Lett.}, vol. 4, no. 1, pp. 10-13, Feb. 2015.
\bibitem{Z. Chu3}
Z. Chu, Z. Zhu, M. Johnston, and S. Le Goff, \lq\lq Simultaneous wireless information power transfer for MISO secrecy channel,\rq\rq \ \emph{IEEE Trans. Vehicular Technol.}, to be published, 2015.
\bibitem{S. H. Wang}
S. H. Wang and B. Y. Wang, \lq\lq Robust secure transmit design in MIMO channels with simultaneous wireless information and power transfer,\rq\rq \ \emph{IEEE Signal Process. Lett.}, vol. 22, no. 11, pp. 2147-2151, Nov. 2015.
\bibitem{F. Wang}
F. Wang, T. Peng, Y. W. Huang, and X. Wang, \lq\lq Robust transceiver optimization for power-splitting based downlink MISO SWIPT systems,\rq\rq \ \emph{IEEE Signal Process. Lett.}, vol. 22, no. 9, pp. 1492-1496, Sept. 2015.
\bibitem{A. Ben-Tal}
A. Ben-Tal, L. El Ghaoui, and A. Nemirovski, \emph{Robust Optimization}. Princeton, NJ, USA: Princeton Univ. Press, 2009.
\bibitem{S. P. Boyd}
S. P. Boyd and L. Vandenberghe, \emph{Convex Optimization}. Cambridge, U.K.: Cambridge Univ. Press, 2004.
\bibitem{M. Grant}
M. Grant and S. Boyd, CVX: Matlab software for disciplined convex programming 2011 [Online]. Available: http://cvxr.com/cvx.
\bibitem{I. Bechar}
I. Bechar, \lq\lq A Bernstein-type inequality for stochastic processes of quadratic forms of Gaussian variables.\rq\rq \ Avaliable: http://arxiv.org/abs/0909.3595.
\bibitem{Z. Q. Luo}
Z. Q. Luo, W. K. Ma, A. M. C. So, Y. Ye, and S. Zhang, \lq\lq Semidefinite relaxation of quadratic optimization problems,\rq\rq \ \emph{IEEE Signal Process. Mag.}, vol. 27, no. 3, pp. 20-34, May 2010.
\bibitem{R. K. Sharma}
R. K. Sharma and D. B. Rawat, \lq\lq Advances on security threats and countermeasures for cognitive radio networks: a survey,\rq\rq \ \emph{IEEE Commun. Surveys Tuts.}, vol. 17, no. 2, pp. 1023-1043, Feb. 2015.
\bibitem{Y. Zou1}
Y. Zou, X. Li, and Y. Liang, \lq\lq Secrecy outage and diversity analysis of cognitive radio systems,\rq\rq \ \emph{IEEE J. Sel. Areas Commun.}, vol. 32, no. 11, pp. 2222-2236, Nov. 2014.
\bibitem{N. Mokari}
N. Mokari, S. Parsaeefard, H. Saeedi, P. Azmi, and E. Hossain, \lq\lq Secure robust ergodic uplink resource allocation in relay-assisted cognitive radio networks,\rq\rq \ \emph{IEEE Trans. Signal Process.}, vol. 63, no. 2, pp. 291-304, Jan. 2015.
\bibitem{A. Ben-Tal1}
A. Ben-Tal and A. Nemirovski, \lq\lq Lectures on modern convex optimization: Analysis, Algorithms, and Engineering Applications,\rq\rq \ in \emph{MPSSIAM Series on Optimization}. Philadelphia, PA, USA: SIAM, 2001.
\end{thebibliography}
\end{document}